\documentclass{emulateapj}

\newcommand{\Msun}{\ensuremath{M_{\odot}}}

\newcommand{\Hb}{\ensuremath{{\rm H}\beta}}

\newcommand{\HdF}{\ensuremath{{\rm H}\delta_{\rm F}}}

\newcommand{\Mgb}{\ensuremath{{\rm Mg}\, b}}
\newcommand{\Fe}{\ensuremath{\langle {\rm Fe}\rangle}}
\newcommand{\MgFe}{\ensuremath{[{\rm MgFe}]^{\prime}}}

\newcommand{\OFe}{\ensuremath{{\rm O}/{\rm Fe}}}
\newcommand{\aFe}{\ensuremath{\alpha/{\rm Fe}}}
\newcommand{\FeH}{\ensuremath{{\rm Fe}/{\rm H}}}
\newcommand{\ZH}{\ensuremath{Z/{\rm H}}}

\newcommand{\Mgtwo}{\ensuremath{{\rm Mg}_2}}

\newcommand{\Age}{Age}

\slugcomment{The Astrophysical Journal}
\received{2004 June, 30}
\accepted{2004 October, 5}
\shorttitle{THE EPOCHS OF EARLY-TYPE GALAXY FORMATION}
\shortauthors{D.\ THOMAS, C.\ MARASTON, R.\ BENDER, C.\ MENDES DE OLIVEIRA}

\begin{document}

\bibliographystyle{apj}

\title{The epochs of early-type galaxy formation as a function of
environment}

\author{Daniel Thomas, Claudia Maraston, Ralf Bender} 
\affil{Max-Planck-Institut f\"ur extraterrestrische Physik,
Giessenbachstra\ss e, D-85748 Garching, Germany\\
Universit\"ats-Sternwarte M\"unchen, Scheinerstr.~1, D-81679
M\"unchen, Germany}
\and
\author{Claudia Mendes de Oliveira}
\affil{Instituto Astron\^omico e Geof\'{\i}sico, Universidade de S\~ao
   Paulo, Rua do Mat\~ao 1226 - Cidade Universit\'aria 05508-900 S\~ao
   Paulo SP - Brazil}

\begin{abstract}
The aim of this paper is to set constraints of the epochs of
early-type galaxy formation through the 'archaeology' of the stellar
populations in local galaxies.  Using our models of absorption line
indices that account for variable abundance ratios, we derive ages,
total metallicities, and element ratios of 124 early-type galaxies in
high and low density environments.  The data are analyzed by
comparison with mock galaxy samples created through Monte Carlo
simulations taking the typical average observational errors into
account, in order to eliminate artifacts caused by correlated errors.
We find that all three parameters age, metallicity, and \aFe\ ratio
are correlated with velocity dispersion. We show that these results
are robust against recent revisions of the local abundance pattern at
high metallicities. To recover the observed scatter we need to assume
an intrinsic scatter of about 20 per cent in age, $0.08\;$dex in [\ZH]
and $0.05\;$dex in [\aFe].  All low-mass objects with $M_*\la
10^{10}\;\Msun$ ($\sigma\la 130\;$km/s) show evidence for the presence
of intermediate-age stellar populations with low \aFe\ ratios.  About
20 per cent of the intermediate-mass objects with $10^{10}\la
M_*/\Msun\la 10^{11}$ ($110\la\sigma/{\rm km/s}\la 230$, both
ellipticals and lenticulars) must have either a young subpopulation or
a blue horizontal branch.
Based on the above relationships valid for the bulk of the sample, we
show that the Mg-$\sigma$ relation is mainly driven by metallicity,
with similar shares from the \aFe\ ratio ($23$ per cent) and age ($17$
per cent).  We further find evidence for an influence of the
environment on the stellar population properties. Massive early-type
galaxies in low-density environments appear on average $\sim 2\;$Gyrs
younger and slightly ($\sim 0.05-0.1$~dex) more metal-rich than their
counterparts in high density environments. No offsets in the \aFe\
ratios, instead, are detected. With the aid of a simple chemical
evolution model, we translate the derived ages and \aFe\ ratios into
star formation histories. We show that most star formation activity in
early-type galaxies is expected to have happened between redshifts
$\sim 3$ and 5 in high density and between redshifts 1 and 2 in low
density environments. 
We conclude that at least 50 per cent of the total stellar mass
density must have already formed at $z\sim 1$, in good agreement with
observational estimates of the total stellar mass density as a
function of redshift.  Our results suggest that significant mass
growth in the early-type galaxy population below $z\sim 1$ must be
restricted to less massive objects, and a significant increase of the
stellar mass density between redshifts 1 and 2 should be present
caused mainly by the field galaxy population.  The results of this
paper further imply vigorous star formation episodes in massive
objects at $z\sim 2$--$5$ and the presence of evolved ellipticals
around $z\sim 1$, both observationally identified as SCUBA galaxies
and EROs, respectively.
\end{abstract}



\section{Introduction}
\label{intro}
The most direct way to constrain the formation and evolution of
galaxies certainly is to trace back their evolution with redshift
\citep[e.g.][]{Araetal93,BZB96,ZB97,SED98,KBB99,Pogetal99,Zieetal99,Keletal00,vDoketal00,Sagetal00}.
The price to be paid, however, is that high-redshift data naturally
have lower quality and are therefore more difficult to interpret.  A
clear complication is the so-called progenitor bias, which implies
that galaxies observed at low and high redshift are not necessarily
drawn from the same sample \citep{vDoketal00}.  The alternative
approach is the detailed investigation of the stellar populations in
local galaxies, which has been pioneered by analyzing slopes and
scatter of color-magnitude and scaling relations of early-type
galaxies \citep[e.g.][]{Dreetal87,DD87,BLE92,BBF92,BBF93,RC93},
followed by a number of detailed studies of absorption line indices
(e.g., \citealt{P89}; \citealt*{GAS90}; \citealt*{WFG92};
\citealt*{DSP93}; \citealt{CD94}; \citealt{Rosetal94}; \citealt{BP95};
\citealt*{FFI95}; \citealt*{JFK95}; \citealt{Greggio97};
\citealt{Vazetal97}; \citealt*{TCB98}; \citealt{KD98};
\citealt{Mehetal98}; \citealt{Worthey98}; \citealt{KA99};
\citealt{VA99}; \citealt{Jor99}; \citealt{Teretal99}; \citealt{Kun00};
\citealt{Kunetal01}; \citealt{Lonetal00}; \citealt{Traetal00b};
\citealt{Pogetal01a,Pogetal01b}; \citealt{Davetal01};
\citealt{Caretal02}; \citealt{PS02}; \citealt{TF02};
\citealt{Sagetal02}; \citealt{Cenetal03}; \citealt*{CRC03};
\citealt{Mehetal03}; \citealt*{TMB03b}; and others).  We call this the
'archaeology approach'.  The confrontation with predictions from
models of galaxy formation is certainly most meaningful, when the two
approaches, the mining of the high-redshift universe and the
archaeology of local galaxies set consistent constraints. In this
paper we follow the latter approach.

The main challenge in the archaeology of stellar population is the
disentanglement of age and metallicity effects. The use of absorption
line indices to lift this degeneracy is powerful
\citep{Fabetal85,G93,Worthey94}, but has been up to now hampered by
the fact that different metallic line indices yield different
metallicities and therefore different ages (see references above).  We
have solved this problem by developing stellar population models that
include element abundance ratio effects and now allow for an
un-ambiguous derivation of age, total metallicity, and element ratios
from (Lick) absorption line indices \citep*[][hereafter TMB]{TMB03a}.
A complication that remains, however, is the degeneracy between age
and horizontal branch morphology, which stems from the fact that the
presence of warm horizontal branch stars (which cannot be excluded)
strengthens the Balmer absorption and can mimic a younger stellar
population age \citep{FB95,MT00,Leeetal00}.  We will additionally
discuss this problem in the present paper.

The principal aim of this study is to constrain the formation epochs
of the stellar populations in early-type galaxies as a function of
their type (elliptical and lenticular), mass and environmental
density.  For this purpose we analyze a homogeneous, high-quality data
sample of 124 early-type galaxies in various environmental densities.
With our new stellar population models (TMB) we determine ages,
total metallicities, and \aFe\ ratios from the absorption line indices
\Hb, \Mgb, and \Fe, and seek for possible correlations of these
parameters with velocity dispersion.  First tentative results have
already been discussed in \citet*{TMB02b}. Here we present the final
and more comprehensive analysis. In particular, we compare the
observed data set with mock samples produced through Monte Carlo
simulations taking the typical average observational errors into
account, in order to eliminate the confusion caused by correlated
errors \citep{Traetal00b,Kunetal01,TF02}.

The \aFe\ element ratio plays a key role for the accomplishment of our
main goal, namely the derivation of formation epochs, While the
so-called $\alpha$-elements O, Ne, Mg, Si, S, Ar, Ca, Ti (particles
that are build up with $\alpha$-particle nuclei) plus the elements N
and Na are delivered mainly by Type~II supernova explosions of massive
progenitor stars, a substantial fraction of the Fe-peak elements Fe
and Cr comes from the delayed exploding Type~Ia supernovae
\citep*[e.g.][]{NTY84,WW95,TNH96}. Hence, the \aFe\ ratio quantifies
the relative importance of Type~II and Type~Ia supernovae
\citep*{GR83,MG86,PT95,TGB98}, and therefore carries information about
the timescale over which star formation occurs.  Thus, the \aFe\ ratio
can be considered as an additional measure of late star formation, and
we will use it both to constrain formation timescales and to lift the
degeneracy between age and horizontal branch morphology.

The paper is organized as follows. In Sections~\ref{sec:datasample}
and~\ref{sec:montecarlo} we present the methodology of the paper,
i.e.\ we introduce the data sample, we briefly summarize the stellar
population model of TMB, and we explain the construction of the mock
galaxy sample through Monte Carlo simulations. The results of this
analysis are presented in Section~\ref{sec:results}. We use these
results to explore the origin of the Mg-$\sigma$ relation
(Section~\ref{sec:mgsigma}) and to derive star formation histories
(Section~\ref{sec:sfh}) with the aim to constrain the epoch of
early-type galaxy formation. The main results of the paper are
discussed and summarized in Sections~\ref{sec:discussion}
and~\ref{sec:conclusions}.


\section{The data sample}
\label{sec:datasample}
\subsection{Sample selection}
\begin{figure}
\centering\includegraphics[width=\linewidth]{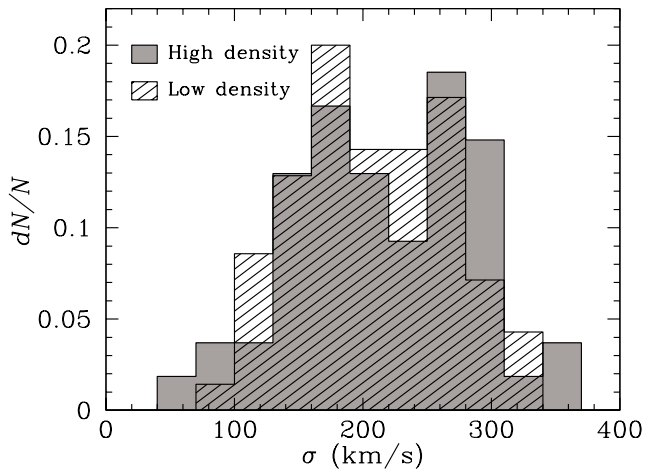}
\caption{Distribution of central velocity dispersion in high density
(grey shaded histogram) and low density (hatched histogram)
environments.}
\label{fig:histsig}
\end{figure}

We analyze a sample of 124 early-type galaxies, 70 of which reside in
low-density and 54 in high-density environments, containing roughly
equal fractions of elliptical and lenticular (S0) galaxies.  To
maintain statistical significance, in the following we make the rough
bimodal distinction between low-density and high-density
environment. It should be emphasized that the classification 'low
density' comprises galaxies in environments with low density, not
necessarily isolated objects.  The sample is constructed from the
following sources: 41 galaxies in high (Virgo cluster) and low density
environments \citep{G93}, 32 Coma cluster galaxies
\citep{Mehetal00,Mehetal03}, and 51 galaxies (mostly low density
environment) from \citet{Beuetal02} selected from the ESO--LV catalog
\citep{LV89} (highest-quality objects).  In this latter sample,
objects with a local galaxy surface density NG$_{\mathrm{T}}>9$ are
assigned to the high density environment. NG$_{\mathrm{T}}$ is given
in \citet{LV89} and is the number of galaxies per square degree inside
a radius of one degree around the considered galaxy
\citep[see][]{Beuetal02}. The sample spans a large range in central
velocity dispersion $60\la \sigma_0/{\rm kms}^{-1}\la 340$ with
similar distributions in both environments (Fig.~\ref{fig:histsig}).

In the data samples quoted above, absorption line strengths are
measured as functions of galaxy radius.  We adopt the central indices
determined within 1/10 of the effective radius, so that the analysis
presented here does not suffer from aperture effects. We use the Lick
indices \Hb, \Mgb, and $\Fe=0.5({\rm Fe5270}+{\rm Fe5335})$ to derive
the stellar population parameters age $t$, total metallicity [\ZH],
and the element abundance ratio [\aFe].  The medians of the 1-$\sigma$
errors in \Hb, \Mgb, Fe5270, and Fe5335 are 0.06, 0.06, 0.07,
0.08$\;$\AA, respectively.

We have re-observed 19 objects of the \citet{Beuetal02} sample with
the same telescope (the 1.5m telescope on La Silla) with the principal
aim of extending the wavelength range to the blue (Mendes de Oliveira
et al., in preparation). We find a good agreement for the optical
absorption line indices \Hb, \Mgb, and \Fe\ between the two samples,
with a small systematic offset in \Mgb\ of $0.2\;$\AA.  As the latter
observations have higher quality due to the installation of a new CCD
chip and holographic grating, we corrected for the offset in \Mgb\ and
replaced the values of \citet{Beuetal02} with the new measurements for
the objects in common. We emphasize here, however, that the results of
this paper are not affected by this procedure.

\subsection{Absorption line indices}
\begin{figure*}
\centering\includegraphics[width=0.7\linewidth]{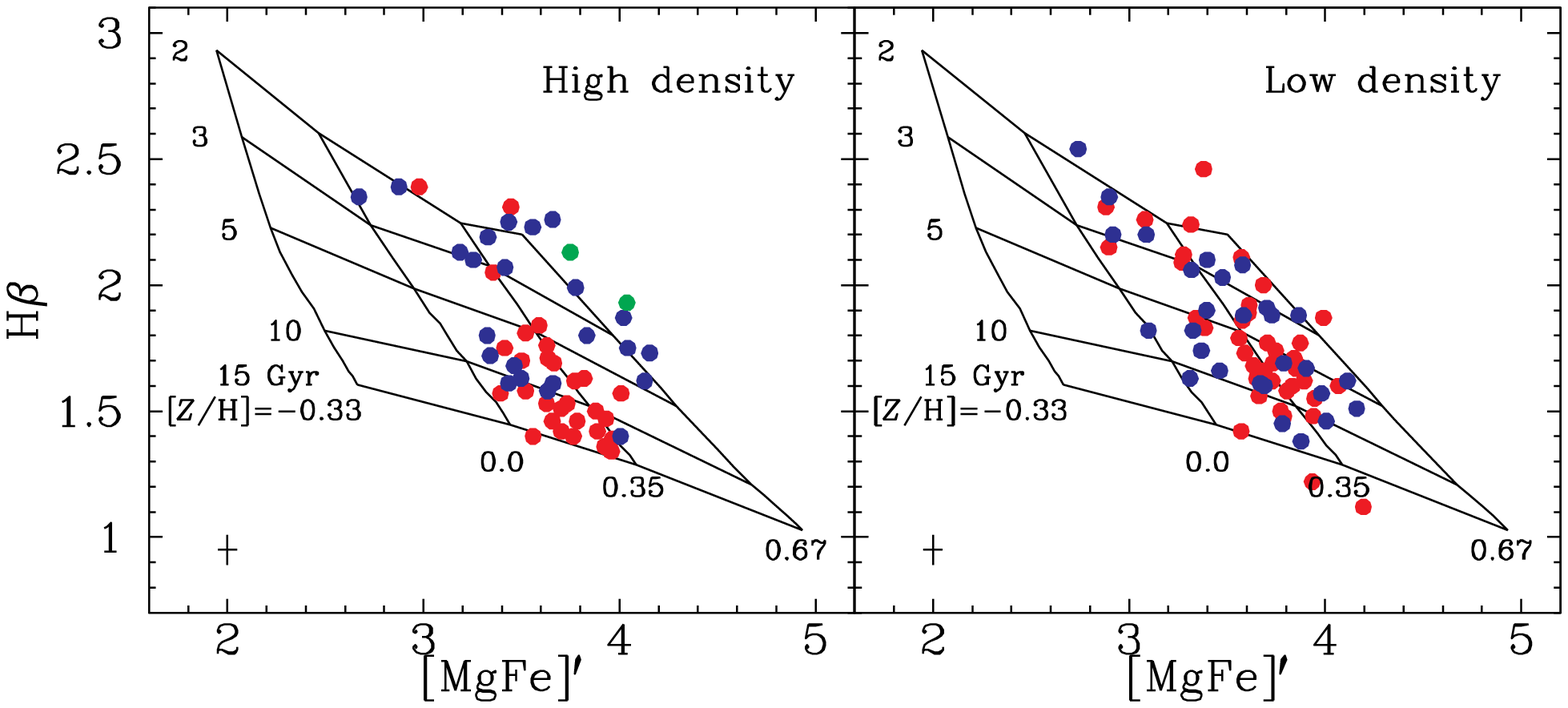}
\centering\includegraphics[width=0.7\linewidth]{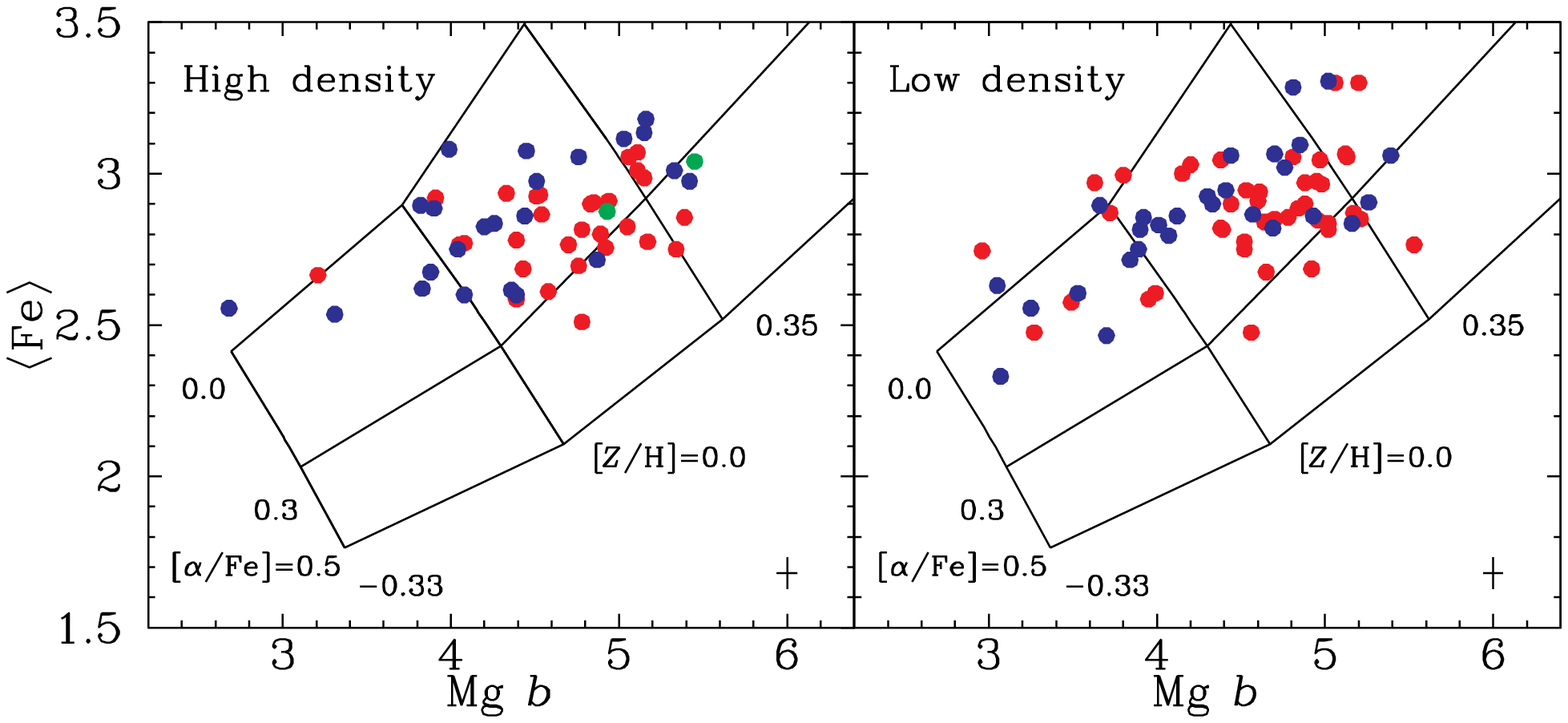}
\caption{Lick indices \MgFe\ versus \Hb\ (top panels) and \Mgb\ versus
\Fe\ (bottom panels). Symbols are the early-type galaxies analyzed in
this paper, taken from \citet{G93}, \citet{Beuetal02}, and
\citet{Mehetal03}.  Index values are measured within $\sim 1/10\ r_e$.
Left and right panels show objects in high and low density
environments, respectively. Red symbols are elliptical, blue symbols
lenticular, and green symbols are cD galaxies. Median 1--$\sigma$
error bars are shown. SSP models (TMB) with the metallicities
$[\ZH]=0.0,0.35,0.67$ are plotted for different ages
$t=2,3,5,10,15$~Gyr at fixed \aFe\ ratio ($[\aFe]=0$) (top panels) and
different \aFe\ ratios $[\aFe]=0.0,\ 0.3,\ 0.5$ at fixed age
($t=12$~Gyr) (bottom panels) as indicated by the labels.}
\label{fig:datagrid}
\end{figure*}
In Fig.~\ref{fig:datagrid} we present the absorption line index
measurements plotting \MgFe\ vs.\ \Hb\ (top panels) and \Mgb\ vs.\
\Fe\ (bottom panels).  The index 
\[
\MgFe=\sqrt{\Mgb\cdot (0.72\cdot {\rm Fe5270}+0.28\cdot {\rm Fe5335})}\ ,
\]
a slight modification of [MgFe] defined by \citet{G93}, is almost
completely independent of \aFe\ ratio variations (TMB).  Left and
right panels show data in high density and low density environments,
respectively. Red symbols are elliptical, blue symbols are lenticular,
and green symbols are cD galaxies. Overplotted are our stellar
population models (TMB) for various ages, metallicities, and \aFe\
ratios as indicated by the labels in the diagrams.

\subsubsection{TMB models}
\label{sec:tmb03}
The TMB models take the effects from element abundance ratio changes
on all Lick indices into account, hence give Lick indices of simple
stellar populations (SSPs) not only as functions of age and
metallicity but also as a function of the \aFe\ ratio. They are based
on the evolutionary population synthesis code of \citet{Ma98,Ma05}. In
\citet{TMB03a}, the impact from element ratio changes is computed with
the help of the \citet{TB95} response functions, using an extension of
the method introduced by \citet{Traetal00a}. The updated version used
in this paper adopts the new metallicity-dependent response functions
from \citet*{KMT05}. As discussed there, the impact of metallicity on
the response functions for the indices considered here is negligible
\citep*[see also][]{TMK04}, hence does not affect the results of this
paper.

Particular care was taken to calibrate the SSP models with globular
cluster data \citep{Maretal03}, which---most importantly---include
objects with relatively high, namely solar, metallicities
\citep{Puzetal02}. We match very well their \Hb, \Mgb, and \Fe\
indices with models in which the \aFe\ ratio is enhanced by a factor
two relative to the solar value, in agreement with results from
high-resolution spectroscopy of individual stars (see TMB and
\citealt{Maretal03}, and references therein). Because of the inclusion
of element ratio effects, the models allow for the clear distinction
between total metallicity [\ZH] and the $\alpha$ to iron-peak elements
ratio [\aFe].

We use these models to derive the three stellar population parameters
age, metallicity, and \aFe\ ratio from the three line indices \Hb,
\Mgb, and \Fe\ in a twofold iterative procedure explained below.  As
mentioned in the Introduction, there is a degeneracy between age and
horizontal branch morphology.  A way to lift this degeneracy is the
consideration of the \aFe\ ratio.  We expect young stellar populations
generally to be characterized by lower \aFe\ ratios owing to the late
enrichment of Fe from Type~Ia supernovae.

In this paper we present a detailed investigation of the data set
based on Monte Carlo simulations as shall be discussed in the
following sections. However, first tentative conclusions can already
be drawn by the direct comparison between the observational data and
SSP model grids, as discussed in the following.

\subsubsection{Ages and metallicities}
The index \MgFe\ is a good tracer of total metallicity (still
degenerate with age, of course), as it is almost completely
independent of \aFe\ ratio variations (TMB). As also \Hb\ is only very
little sensitive to \aFe\ (TMB), the top panels of
Fig.~\ref{fig:datagrid} are well suited to read off ages and total
metallicities independent of abundance ratio effects. 

The sample splits in two subclasses divided by the Balmer absorption
index at $\Hb\approx 2\;$\AA. The main part of the data is centered
around $\Hb=1.6\;$\AA, while about one quarter of the objects have
relatively high \Hb\ absorption implying the presence of either young
stellar populations or blue horizontal branch morphologies
\citep{MT00}.  This surprisingly well-defined separation is also
discussed in \citet{Mehetal03} for the Coma cluster objects alone.
Interestingly, the pattern is still present for the larger galaxy
sample considered here, and in particular the fraction of high \Hb\
objects seems to be independent of the environmental density
(Fig.~\ref{fig:datagrid}), but we do not find a significant difference
with respect to the galaxy type. Fig.~\ref{fig:datagrid} further hints
toward a difference between objects as a function of environment
within the old (low \Hb) class, in the sense that galaxies in low
density environments seem to have on average slightly younger ages and
higher metallicities. This difference is more clearly visible in the
direct comparison of the derived ages discussed below.

\subsubsection{Element abundance ratios}
The bottom panel of Fig.~\ref{fig:datagrid} can be used to get rough
estimates for the \aFe\ ratios of the sample galaxies. It should be
emphasized, however, that the \Mgb-\Fe\ plane is not completely
degenerate in age, and the ages derived from the top panels have to be
taken into account for the precise derivation of \aFe\ ratios. Still
we can infer from the bottom panels of Fig.~\ref{fig:datagrid} that
the \aFe\ ratios range roughly between solar and two times solar, and
that objects with strong \Mgb\ indices, hence velocity dispersion
because of the Mg-$\sigma$ relation \citep[e.g.][]{BBF93}, tend to
have higher \aFe\ ratios. At least no obvious impact from
environmental density on the \aFe\ ratios is visible.

\subsubsection{Derivation of stellar parameters}
The stellar population parameters are derived in a two-fold iterative
procedure.  First, we arbitrarily fix the \aFe\ ratio, and determine
ages and metallicities for the index pairs (\Hb,\Mgb) and (\Hb,\Fe),
by starting with arbitrary age-metallicity pairs, which we modify
iteratively until both index pairs are reproduced. The two
metallicities obtained from \Mgb\ and \Fe, respectively, are used to
adjust the \aFe\ ratio, and to start a new iteration. These steps are
repeated until the age-metallicity pairs derived from (\Hb,\Mgb) and
(\Hb,\Fe) at a given \aFe\ ratio are consistent within 1 per cent
accuracy. For ages and metallicities between the grid points quoted
above, we interpolate linearly.

The resulting ages, metallicities, and \aFe\ for the sample under
investigation are listed in Table~\ref{tab:results}.

\section{The Monte Carlo approach}
\label{sec:montecarlo}
An uncomfortable effect of the degeneracy between age and metallicity
is that the errors of these stellar population parameters are not
independent. It is still a matter of debate, for instance, whether the
age-metallicity anti-correlation found for early-type galaxies in a
number of studies is an artifact caused by correlated errors
\citep{Jor99,Traetal00b,Pogetal01b,Kunetal01,TF02,PS02}. A possibility
to get a handle on these effects is the performance of Monte Carlo
simulations taking into account the observational uncertainties of the
line indices from which the stellar population parameters are
derived. \citet{Kunetal01} follow this method in order to assess the
relationship between ages and metallicities of 72 early-type galaxies
by comparison with a mock sample produced through Monte Carlo
simulations.

\begin{figure}
\includegraphics[width=\linewidth]{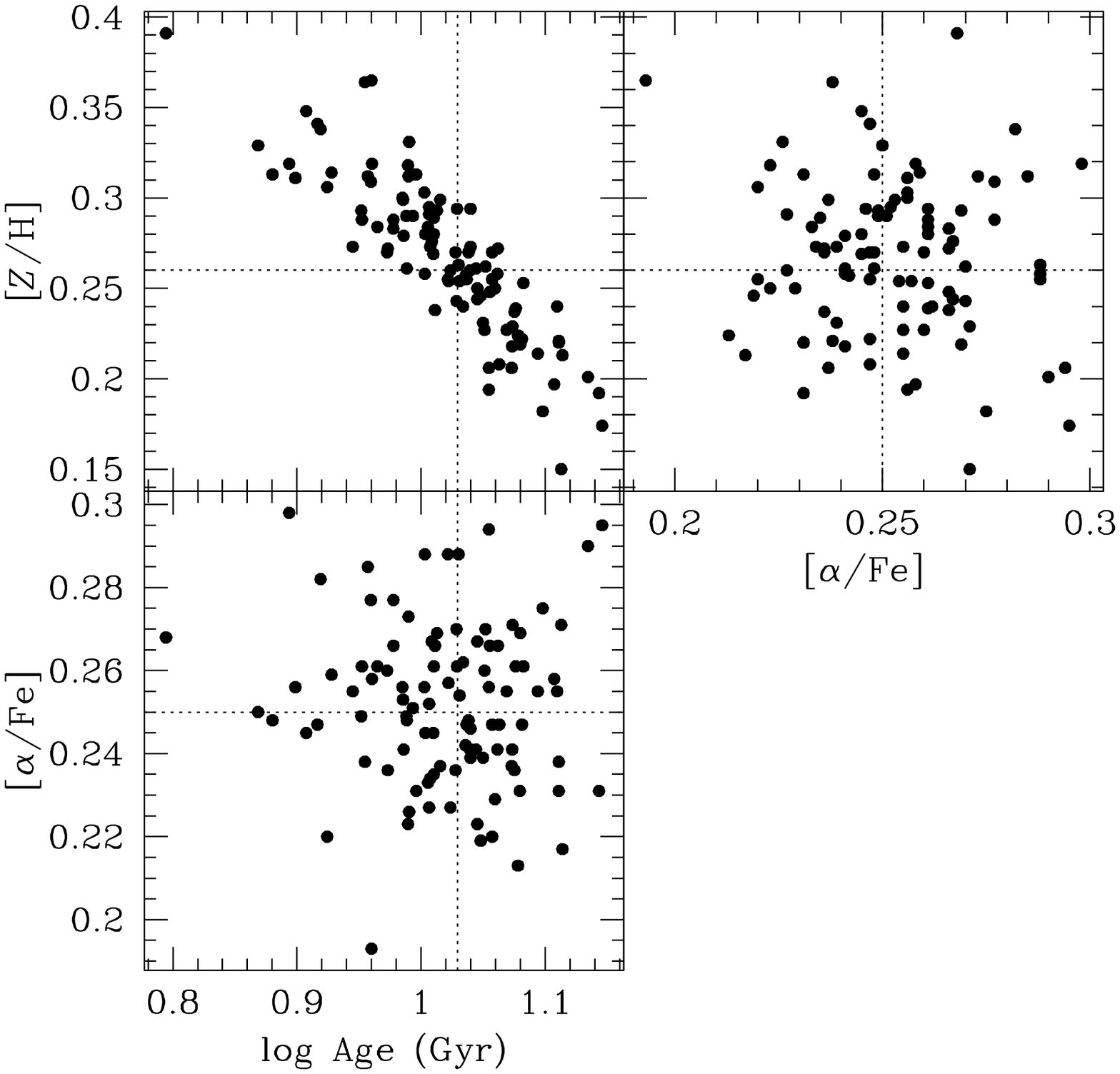}
\caption{Stellar population parameters of an object with
$\Hb=1.59\;$\AA, $\Mgb=4.73\;$\AA, and $\Fe=2.84\;$\AA corresponding
to an age of 10.7$\;$Gyr, a metallicity $[\ZH]=0.26$, and
$[\aFe]=0.25$ as indicated by the dotted lines. The circles are Monte
Carlo realizations perturbing the original index values with the
1-$\sigma$ errors $d\Hb=0.06\;$\AA, $d\Mgb=0.06\;$\AA, $d{\rm
Fe5270}=0.07\;$\AA, and $d{\rm Fe5335}=0.08\;$\AA).  The resulting
1-$\sigma$ errors of the stellar population parameters are $1.48\;$Gyr
in age, $0.04\;$dex in total metallicity, and $0.02\;$dex in \aFe\
ratio.}
\label{fig:monte}
\end{figure}

In this paper we extend the approach of \citet{Kunetal01} to the full
parameter space including also \aFe\ ratios besides age and total
metallicity. We produce a sample of artificial galaxies with given
ages, metallicities, and \aFe\ ratios, from which we compute with the
TMB SSP models the line indices \Hb, \Mgb, and \Fe. By means of Monte
Carlo simulations we perturb these 'exact' values with the 1-$\sigma$
errors quoted above, assuming a Gaussian error distribution. An
illustrative example for the resulting correlations between the errors
of the three population parameters is given in Fig.~\ref{fig:monte}
for an artificial object with $\Hb=1.59\;$\AA, $\Mgb=4.73\;$\AA, and
$\Fe=2.84\;$\AA\ corresponding to an age of 10.7$\;$Gyr, a metallicity
$[\ZH]=0.26$, and $[\aFe]=0.25$ indicated by the dotted lines. The
circles are 100 Monte Carlo realizations taking the 1-$\sigma$ errors
in the line indices ($d\Hb=0.06\;$\AA, $d\Mgb=0.06\;$\AA, $d{\rm
Fe5270}=0.07\;$\AA, $d{\rm Fe5335}=0.08\;$\AA) into account.

The distributions of the resulting stellar population parameters can
be well described by Gaussians. The corresponding 1-$\sigma$ errors
are $1.48\;$Gyr in age, $0.04\;$dex in total metallicity, and
$0.02\;$dex in \aFe\ ratio. We confirm that errors produce a very
tight linear anti-correlation between log~age and metallicity (top
left panel) as already mentioned above.  Interestingly, we find that
also the errors of age and \aFe\ ratio are anti-correlated (bottom
left panel). This relationship is the result of the slightly stronger
age dependence of the \Mgb\ index with respect to \Fe, caused by the
larger contribution from turnoff stars to \Mgb\ and \Mgtwo\
\citep{Maretal03}. An underestimation of the age leads to the
prediction of lower Mg indices relative to iron, which results in the
derivation of a larger \aFe\ ratio, and vice versa.

In the following Monte Carlo simulations will be used to reproduce and
analyze the distribution of the data in the diagrams shown in
Fig.~\ref{fig:datagrid}. The simulations are carried out in two steps,
low density and high density environments are considered separately
throughout the analysis.

1) First, we seek correlations between galaxy velocity dispersion and
the stellar population properties age, metallicity, and \aFe\ ratio
that reproduce the majority of the data sample. Indeed,
\citet{Traetal00b} have already shown that velocity dispersion is the
only structural parameter that is found to modulate the stellar
populations of galaxies. As we aim at deriving average properties,
simple stellar populations of single age and single chemical
composition are used. We start with the velocity dispersion ($\sigma$)
distribution of the data sample shown in Fig.~\ref{fig:histsig}. Based
on arbitrarily chosen relationships, we assign to every $\sigma$ an
age, metallicity, and \aFe\ ratio, which we transform into \Hb, \Mgb,
and \Fe. These indices are then perturbed with the observational error
(see above) adopting a Gaussian probability distribution.  In this way
we obtain a mock galaxy sample that is closest to the observational
situation in terms of size, velocity dispersion distribution, and
error perturbation. If necessary, we add an intrinsic scatter in the
stellar population parameters age, metallicity, and \aFe. This step is
repeated, until the majority of the observed data points in the
index-index diagrams \MgFe\ vs.\ \Hb\ and \Mgb\ vs.\ \Fe\ is
reproduced (comparison 'by eye'). The quality of the fit is further
checked with respect to the zero points, slopes, and scatter of the
relationships between the stellar parameters (age, metallicity, \aFe\
ratio) and velocity dispersion. By means of the index-index diagrams
we identify those objects that are not reproduced by the mock data
sample based on the simple relationships introduced in this step.

2) In a second step, we therefore perturb the general simple scaling
laws found in the first step in 2-component models in order to match
those remaining objects. Like before, we adopt the exact $\sigma$
distribution of the outliers and assign age, metallicity and \aFe\ to
every $\sigma$ based on the above final relationships to define the
base population. Then we add a second component with arbitrary stellar
population parameters and determine the index values of this composite
population. These are then perturbed like in the first step and
compared to the observational data. We note that the contribution in
mass of the second component is treated as the forth free parameter,
hence the solutions will not be unique, as we have only three
observables as constraints (\Hb, \Mgb, \Fe).  As discussed in detail
below, we will use plausibility arguments trying to maintain
astrophysical feasibility of the Monte Carlo realizations, in order to
constrain the relative weights of the subcomponents.


\section{Results}
\label{sec:results}
\begin{figure*}
\centering\includegraphics[width=0.7\linewidth]{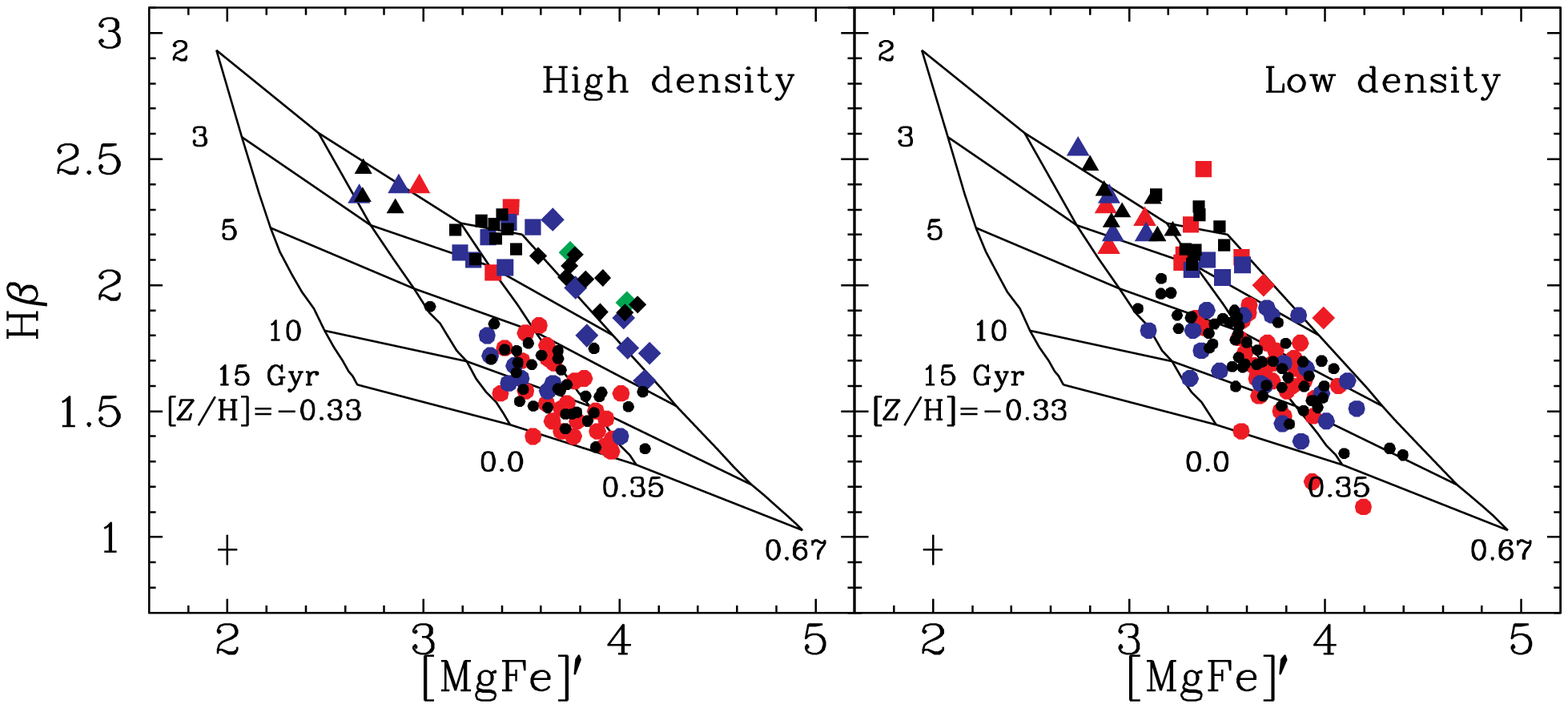}
\centering\includegraphics[width=0.7\linewidth]{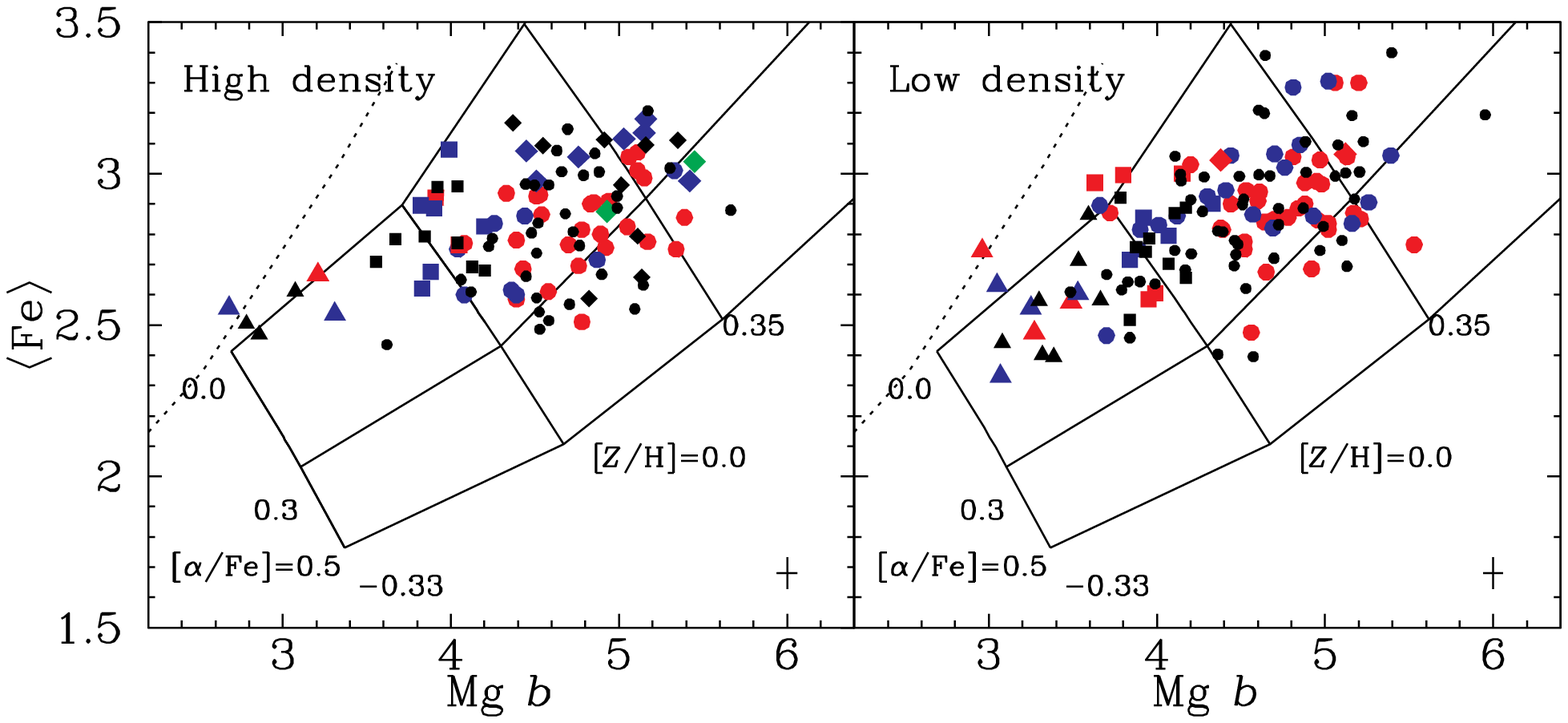}
\caption{Lick indices \MgFe\ versus \Hb\ (top panels) and \Mgb\ versus
\Fe\ (bottom panels).  Colored symbols are the early-type galaxies
analyzed in this paper (red: elliptical, blue: S0, green: cD), taken
from \citet{G93}, \citet{Beuetal02}, and \citet{Mehetal03}, measured
within $1/10\ r_e$.  Median 1--$\sigma$ error bars are shown.  Black
symbols are Monte Carlo realizations. Circles assume linear
relationships of age, metallicity and \aFe\ ratio with velocity
dispersion (see Eqn.~\ref{eqn:relations}). Triangles and squares are
simulations assuming 1.2 and 1.4$\;$Gyr old subcomponents (10 and 20
per cent in mass, respectively) on top of the above relationships,
respectively. Diamonds are realizations in which 70 per cent of the
galaxy's stellar populations is assumed to form around a look back
time of 2.5$\;$Gyr (see Tables~\ref{tab:populations}
and~\ref{tab:sample}).  Left and right panels show objects in high and
low density environments, respectively.  SSP models (TMB) with the
metallicities $[\ZH]=0.0,0.35,0.67$ are plotted for different ages
$t=2,3,5,10,15$~Gyr at fixed \aFe\ ratio ($[\aFe]=0$) (top panels) and
different \aFe\ ratios $[\aFe]=0.0,\ 0.3,\ 0.5$ at fixed age
($t=12$~Gyr) (bottom panels) as indicated by the labels. Dotted lines
in the bottom panels are models for $t=2$~Gyr and $[\aFe]=0$.}
\label{fig:mixgrid}
\end{figure*}
In Fig.~\ref{fig:mixgrid} the comparison of our Monte Carlo
realizations (black symbols) with the observational data (colored
symbols) is shown in the index-index planes introduced with
Fig.~\ref{fig:datagrid}. By means of the 'age-metallicity' diagram
(top panels of Fig.~\ref{fig:mixgrid}) we distinguish between four
different categories, independent of the environment, as indicated by
the symbol types: 1) $V$-light averaged ages older than $\sim
5\;$Gyr (circles), 2) ages between 2 and 3$\;$Gyr and metallicities
$[\ZH]\la 0.35$ (triangles), 3) ages between 2 and 3$\;$Gyr and
metallicities $0.35\la [\ZH]\la 0.67$ (squares), 4) ages between 2 and
5$\;$Gyr and metallicities $[\ZH]\ga 0.67$ (diamonds).

Here we provide a short summary how these categories are reproduced,
detailed descriptions are discussed below in Sections~4.1 to
4.4. Category 1 contains intermediate-mass and massive galaxies
($120\la \sigma/({\rm km/s})\la 320$) and can be reproduced by linear
relationships between the stellar population parameters and
$\log\sigma$. For simplicity we shall call this category {\em old
population}, as Categories 2--4, unlike Category 1, require
2-component models with young sub-populations. Category 2 consists of
{\em low-mass galaxies} ($50\la \sigma/({\rm km/s})\la 130$) and are
reproduced assuming a minor (10 per cent in mass) young ($\sim
1.2\;$Gyr) sub-component. Category 3 are {\em young intermediate-mass}
galaxies ($110\la \sigma/({\rm km/s})\la 230$) with a minor (20 per
cent in mass) young ($\sim 1.4\;$Gyr) component, while Category 4
contains {\em young massive galaxies} ($190\la \sigma/({\rm km/s})\la
360$) with a major (70 per cent in mass) young ($\sim 2.5\;$Gyr)
component. The data sample is well characterized by these four object
classes and well reproduced by the Monte Carlo simulations. In the
following section we will discuss in detail their
characteristics. They are summarized in Table~\ref{tab:populations}
and~\ref{tab:sample}.

\begin{deluxetable}{lcccc}
\tablecaption{Model parameters}
\tablewidth{0pt}
\tablehead{\colhead{} & \colhead{circles} & \colhead{triangles} & \colhead{squares} & \colhead{diamonds} }
\startdata
  $\sigma$ (km/s) & $120$--$320$ & $50$--$130$ & $110$--$230$ &
  $190$--$360$ \\
$\log M_*/\Msun$ & $10$--$12$ & $<10$ & $10$--$11$ & $>11$\\
1st comp. & Eqn.~\ref{eqn:relations} & Eqn.~\ref{eqn:relations}
& Eqn.~\ref{eqn:relations} & Eqn.~\ref{eqn:relations}\\
2nd comp. & no & yes & yes & yes \\
Mass & -- & 10\% & 20\% & 70\% \\
Age (Gyr) & -- & 1.2 & 1.4 & 2.5\\
$[\ZH]$ & -- & $\sim 0.2$ & $\sim 0.8$ & $\sim 0.7$\\
$[\aFe]$ & -- & $\sim -0.06$ & $\sim 0.20$ & $\sim0.34$
\enddata \tablecomments{Stellar masses (row 2) are from
Eqn.~\ref{eqn:masssigma}. The last five rows refer to the second
component.}
\label{tab:populations}
\end{deluxetable}

\subsection{Old population - circles}
\begin{figure}
\includegraphics[width=\linewidth]{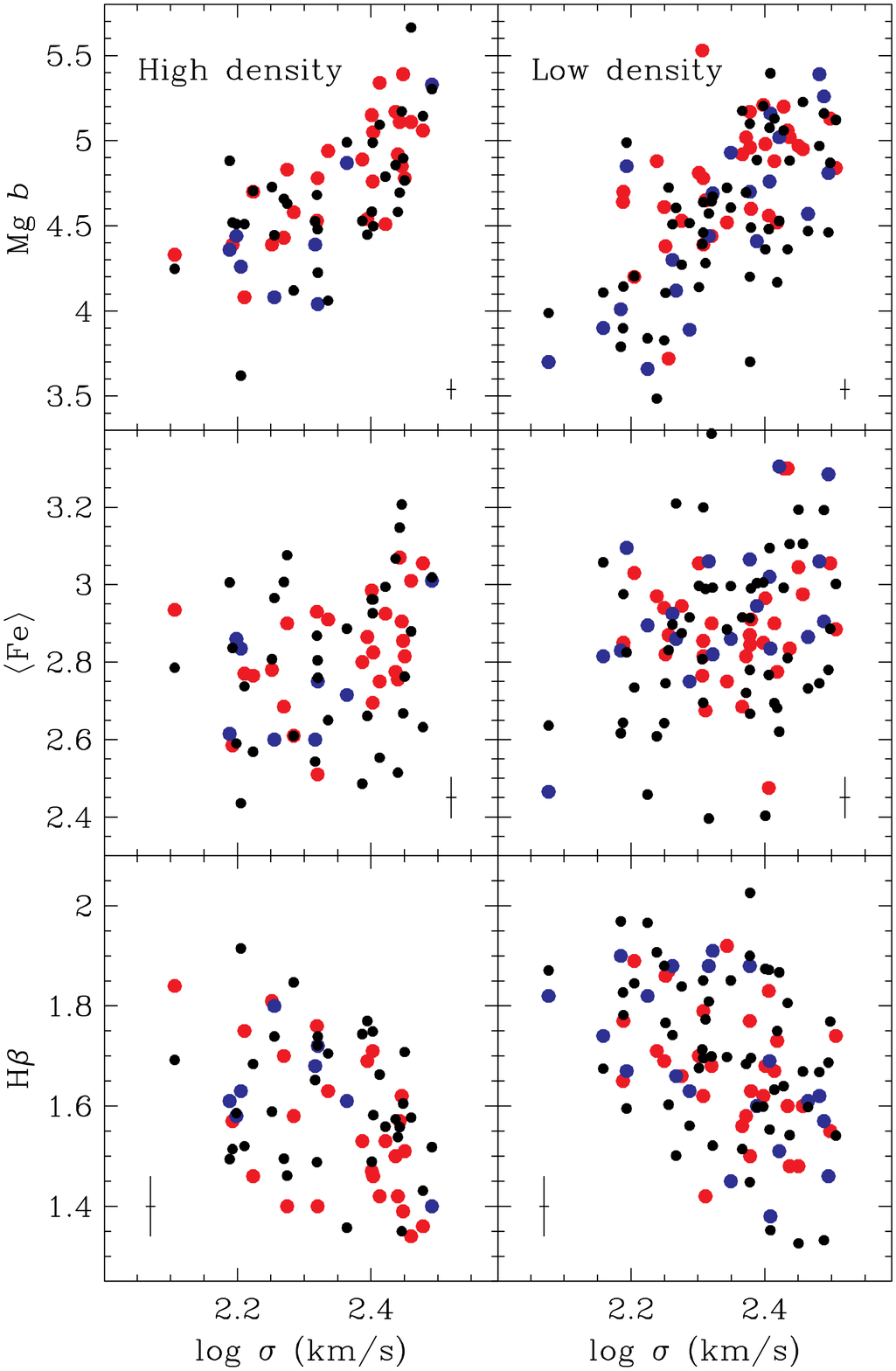}
\caption{Absorption line indices as a functions of velocity dispersion
(measured within $1/10\ r_e$) for the 'old' subpopulation shown as
circles in Fig.~\ref{fig:mixgrid} (red: elliptical, blue: S0, green:
cD, taken from \citet{G93}, \citet{Beuetal02}, and \citet{Mehetal03},
measured within $1/10\ r_e$).  Median 1--$\sigma$ error bars are
shown.  The Monte Carlo realizations are shown by the black
symbols. The latter take the observational errors (see
Fig.~\ref{fig:monte}) into account plus an intrinsic scatter $\Delta
[\aFe]=0.05\;$dex, $\Delta [\ZH]=0.08\;$dex, and $\Delta t=0.2\cdot
t\;$Gyr in high density environments (left panel) and $\Delta
[\aFe]=0.07\;$dex, $\Delta [\ZH]=0.10\;$dex, and $\Delta t=0.25\cdot
t\;$Gyr in low-density environments (right panel).}
\label{fig:indexall}
\end{figure}
This subpopulation represents the major fraction of the present
sample, namely $60$ and $70\;$ per cent in high density and low
density environments, respectively containing both elliptical and
lenticular galaxies. The objects display relatively well-defined
relationships between absorption line indices and central velocity
dispersion, as shown in Fig.~\ref{fig:indexall} by the colored
symbols. Both metallic indices \Mgb\ and \Fe\ correlate with velocity
dispersion, which is well-known for \Mgb\ (e.g., \citealt{BBF93}), and
has also been recently established for \Fe\
\citep{Kunetal01,CRC03,Beretal03d}. The Balmer index \Hb, instead,
clearly tends to decrease with increasing $\sigma$ \citep[see
also][]{Kunetal01,CRC03,Beretal03d}. Fig.~\ref{fig:indexall} shows
that these relationships are essentially independent of the
environment, except that the \Fe-$\sigma$ relation seems somewhat
flatter in low-density environments.

\subsubsection{Scaling relations}
The black symbols in Fig.~\ref{fig:indexall} are the Monte Carlo
realizations corresponding to the dominating 'old' population (circles
in Fig.~\ref{fig:mixgrid}). In these simulations we have assumed
linear correlations of the parameters age, metallicity and \aFe\ ratio
with $\log\sigma$. The dependencies that fit best all three indices as
functions of velocity dispersion for objects in high density
environments are (quantities in brackets are the values derived for
the low-density environment):

\begin{eqnarray}
\label{eqn:relations}
[\aFe] &=& -0.42\ (-0.42) + 0.28\ (0.28)\ \log\sigma\\\nonumber
[\ZH] &=& -1.06\ (-1.03) + 0.55\ (0.57)\ \log\sigma\\\nonumber
\log\ t/{\rm Gyr} &=& 0.46\ (0.17) + 0.238\ (0.32)\ \log\sigma
\end{eqnarray}
The simulations take the observational errors (see
Fig.~\ref{fig:monte}) into account plus an intrinsic scatter $\Delta
[\aFe]=0.05\;$dex, $\Delta [\ZH]=0.08\;$dex, and $\Delta t=0.2\cdot
t\;$Gyr in high density environments and $\Delta [\aFe]=0.07\;$dex,
$\Delta [\ZH]=0.10\;$dex, and $\Delta t=0.25\cdot t\;$Gyr in
low-density environments.  The relationships with velocity dispersion
including their scatter of the line indices are well reproduced. The
relative contributions of the observational errors and the intrinsic
scatter of the stellar population parameters to the overall spread in
the data and simulations will be discussed in more detail below.

Before doing so, we provide estimates of the stellar masses as
a function of velocity dispersion.  For this purpose we transform the
ages and metallicities of Eqn.~\ref{eqn:relations} into stellar
mass-to-light ratios with the help of stellar population models and
use these to derive (via the Faber-Jackson relation, \citealt{FJ76}) a
relationship between stellar mass and velocity dispersion. We adopt
the SSP models of \citet{Ma05} for Kroupa IMF (note that a Salpeter
IMF yields M/L ratios, hence masses, by a factor 1.6 higher). The M/L
ratios (and masses) derived in this way from the high density and low
density relationships vary by only 25 per cent, because the younger
ages of the low density objects are balanced by higher
metallicities. In the following equation we therefore provide the
resulting relationship between stellar mass $M_*$ and $\log\sigma$
averaged over both environments.
\begin{equation}
\label{eqn:masssigma}
\log\ M_* \approx  0.63 + 4.52\ \log\sigma
\end{equation}
This yields the following relations between the stellar population
parameters and galaxy stellar mass.
\begin{eqnarray}
\label{eqn:massrelations}
[\aFe] &=& -0.459\ (-0.459) + 0.062\ (0.062)\ \log\ M_*/\Msun\\\nonumber
[\ZH] &=& -1.137\ (-1.109) + 0.1217\ (0.1261)\ \log\ M_*/\Msun\\\nonumber
\log\ t/{\rm Gyr} &=& 0.427\ (0.125) + 0.053\ (0.071)\ \log\ M_*/\Msun
\end{eqnarray}

\subsubsection{Stellar population parameters}
\begin{figure}
\includegraphics[width=\linewidth]{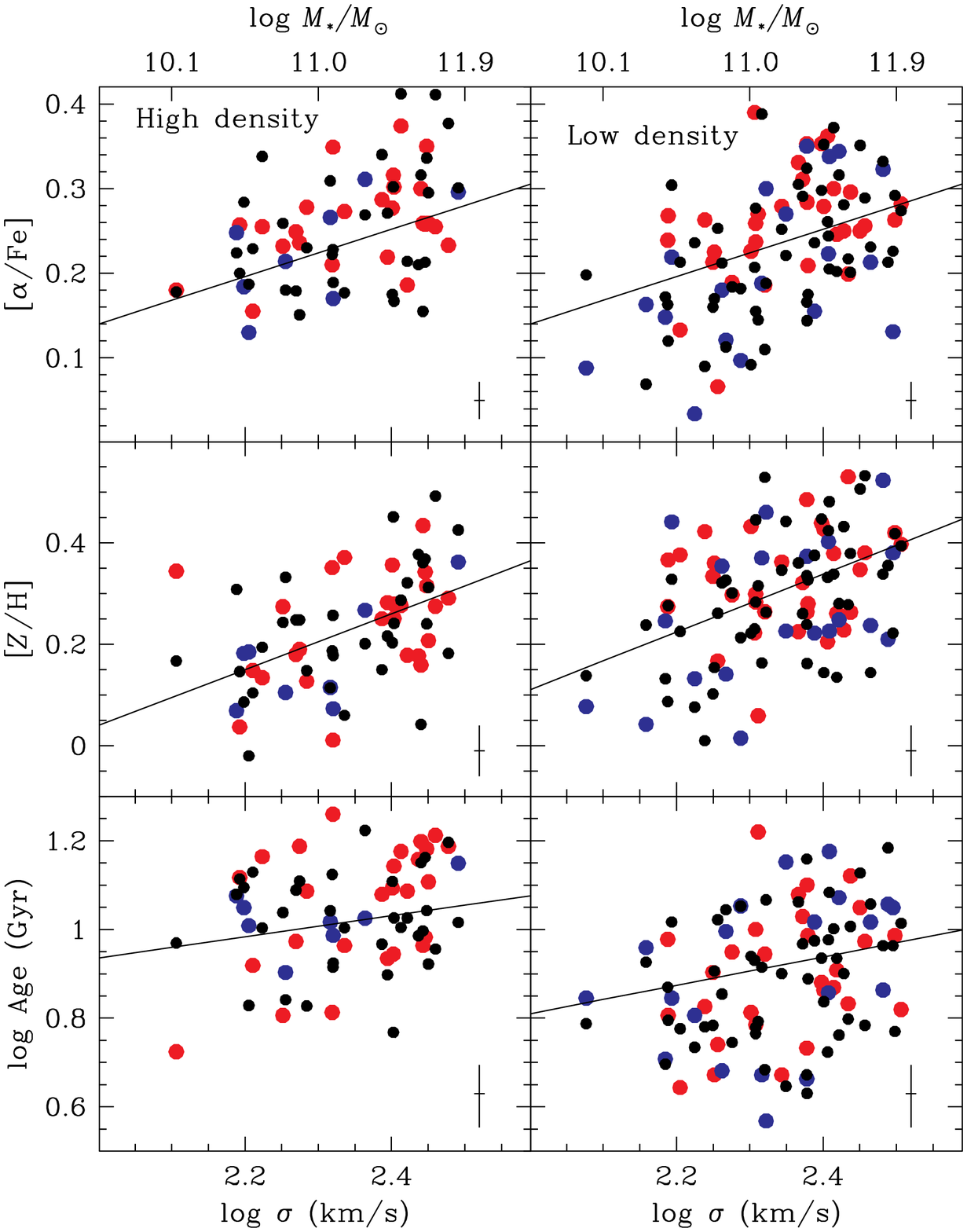}
\caption{Stellar population parameters as a functions of velocity
dispersion (measured within $1/10\ r_e$) and stellar mass (upper
x-axis) for the 'old' subpopulation (circles in Fig.~\ref{fig:mixgrid}
and Fig.~\ref{fig:indexall}).  Ages and element abundances are derived
with the [\aFe] enhanced SSP models described in
Section~\ref{sec:datasample} from the indices \Hb, \Mgb, and
$\Fe=({\rm Fe5270}+{\rm Fe5335})/2$.  Colored symbols are the values
derived from the observational data (red: elliptical, blue: S0, green:
cD), median 1--$\sigma$ error bars are shown.  The Monte Carlo
realizations are shown by the black circles.  The latter take the
observational errors (see Fig.~\ref{fig:monte}) into account plus an
intrinsic scatter $\Delta [\aFe]=0.05\;$dex, $\Delta [\ZH]=0.08\;$dex,
and $\Delta t=0.2\cdot t\;$Gyr in high density environments (left
panel) and $\Delta [\aFe]=0.07\;$dex, $\Delta [\ZH]=0.10\;$dex, and
$\Delta t=0.25\cdot t\;$Gyr in low-density environments (right panel).
Typical error-bars are given in the bottom-right corners.}
\label{fig:popsall}
\end{figure}
\begin{figure*}
\centering\includegraphics*[width=0.8\textwidth]{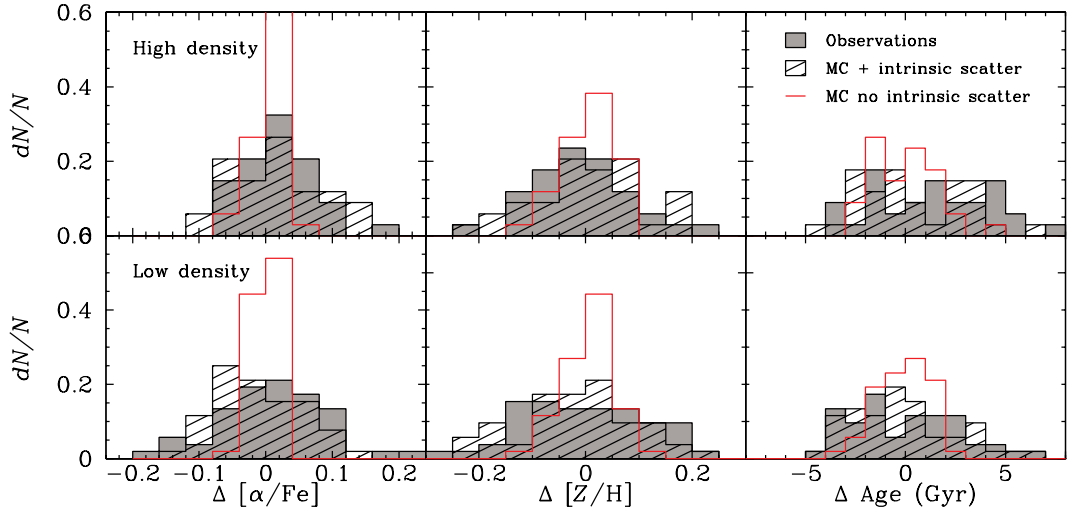}
\caption{Residuals of the relations shown in Fig.~\ref{fig:popsall}
for the observationally derived data (grey histograms) and the
simulations (hatched histograms).  The latter take the observational
errors (see Fig.~\ref{fig:monte}) plus an intrinsic scatter $\Delta
[\aFe]=0.05\;$dex, $\Delta [\ZH]=0.08\;$dex, and $\Delta t=0.2\cdot
t\;$Gyr in high density environments (top panels) and $\Delta
[\aFe]=0.07\;$dex, $\Delta [\ZH]=0.10\;$dex, and $\Delta t=0.25\cdot
t\;$Gyr in low-density environments (bottom panels) into account.
The red histograms show the residuals of simulations without intrinsic
scatter of the stellar population parameters.}
\label{fig:residuals}
\end{figure*}

Fig.~\ref{fig:popsall} shows velocity dispersion and stellar mass
versus the stellar population parameters age, metallicity [\ZH], and
[\aFe] ratio derived from the line indices shown in
Fig.~\ref{fig:indexall}.  They are summarized in
Table~\ref{tab:results}.  Observational data are the colored symbols
(see Fig.~\ref{fig:datagrid}), the Monte Carlo realizations are shown
as black circles. The lines indicate the linear relationships of the
stellar population parameters with $\log\sigma$ used in the
simulations (see Eqn.~\ref{eqn:relations}). The residuals of the data
and Monte Carlo simulations from these relationships are shown in
Fig.~\ref{fig:residuals} by the grey and hatched histograms,
respectively. Note that the simulations take both the observational
errors plus an intrinsic scatter of the stellar population parameters
at a given $\sigma$ ($\Delta [\aFe]=0.05\;$dex, $\Delta
[\ZH]=0.08\;$dex, and $\Delta t=0.2\cdot t\;$Gyr in high density
environments and $\Delta [\aFe]=0.07\;$dex, $\Delta [\ZH]=0.10\;$dex,
and $\Delta t=0.25\cdot t\;$Gyr in low-density environments) into
account. The red histograms are simulations without intrinsic scatter
for comparison.

\medskip
\paragraph*{\bf\boldmath\aFe}
We find a very clear correlation between \aFe\ ratio and $\log\sigma$
or stellar mass. The existence of this relation has already been
anticipated qualitatively by a number of authors
\citep[e.g.][]{WFG92,FFI95,Greggio97,Jor99,Kun00,TF02}, and has only
recently been confirmed quantitatively by means of stellar population
models with variable element abundance ratios
\citep{Traetal00b,PS02,TMB02b,Mehetal03}.  The observed scatter about
the relation can only be reproduced assuming an intrinsic scatter in
[\aFe] ratio of the order $0.05\;$dex as shown by the hatched
histograms in Fig.~\ref{fig:residuals}. In the simulations without
intrinsic scatter (red histograms in Fig.~\ref{fig:residuals}), the
scatter in \aFe\ is significantly smaller than the observed
distribution.

Most interestingly, as already shown in \citet{TMB02b}, the
\aFe-$\sigma$ relation (and its scatter) of early-type galaxies is
independent of the environmental density (see also
Eqn.~\ref{eqn:relations}). This reinforces and extends the results of
the studies by \citet*{JFK95} and \citet{TF02}, which are based on
stellar population models tied to solar element ratios.  Curiously,
early-type galaxies in low density environments seem to exhibit
stronger CN absorption features than in high density environments
\citep{Sanetal03}. As a significant fraction of C and N comes from
intermediate-mass stars in the mass range $5\la M/\Msun\la 8$
\citep{RV81}, this points toward about $10^8$ years longer formation
timescales in low density environments.  A delay well above this value
can certainly be excluded, as the \aFe\ ratio is independent of the
environment. Hence, if true the above result means that there must be
an amazingly high degree of fine tuning in the formation timescales of
early-type galaxies as a function of environmental density.

\medskip
\paragraph*{\bf\boldmath\ZH}
Besides \aFe, also total metallicity \ZH\ correlates with $\sigma$,
hence more massive early-type galaxies are also more metal-rich
\citep[see also][]{Greggio97}. This result is sensible in terms of
chemical evolution, as the deeper potential well of massive objects
hampers the development of a galactic wind, which leads to a more
complete chemical processing to higher element abundances
\citep[e.g.][]{AY87,Ma94,Ed90}.  The observational scatter is best
reproduced assuming about 0.1$\;$dex of intrinsic scatter in total
metallicity at a given velocity dispersion (hatched histograms in
Fig.~\ref{fig:residuals}). Different from the \aFe\ ratio, however,
simulations without intrinsic scatter still provide a reasonable
representation of the data (red histograms in
Fig.~\ref{fig:residuals}).

Unlike for the \aFe\ ratio, we detect a slight impact of the
environmental density in the sense that galaxies in low density
environments appear about 0.05--0.1$\;$dex more metal-rich than
their counterparts in high density environments (see also
Eqn.~\ref{eqn:relations}). This implies that galaxies in high-density
environments seem to be more efficient in ejecting interstellar
material in galactic winds, which might be caused by galaxy
interactions and/or gas stripping processes.

\medskip
\paragraph*{\bf\Age}
As noted in previous studies \citep{Traetal00b,Kunetal01,TF02}, age
seems to be the weakest variable among early-type galaxies, the data
show no significant trend between age and velocity
dispersion. However, it has also been shown by \citet{Kunetal01} that
correlated errors of age and metallicity tend to dilute a correlation
between age and metallicity, because a slight overestimation of \ZH\
leads to an underestimation of the age and vice versa. In
Fig.~\ref{fig:popsall} we show that the observational data are best
reproduced by a relatively flat but significant correlation between
age and velocity dispersion (see solid lines and
Eqn.~\ref{eqn:relations}) in agreement with previous studies
\citep*{PS02,Proetal04,PFB04}.  The scatter of the observationally
derived age (and of the Balmer index \Hb, see Fig.~\ref{fig:indexall})
is relatively well reproduced by the simulations without any intrinsic
scatter (red histograms in Fig.~\ref{fig:residuals}).  A scatter of
the order of 20 per cent at a given $\sigma$ improves the fit as shown
by the hatched histograms in Fig.~\ref{fig:residuals}.

To conclude, the mean age of early-type galaxies is weakly correlated
with galaxy mass with about 20 per cent intrinsic scatter. From this
anti-hierarchical relationship we infer that the stellar populations
of more massive early-type galaxies have formed earlier, which is
further supported by their high \aFe\ ratios.

Unlike in the case of the \aFe\ ratio and total metallicity, we find a
significant difference between the ages of galaxies in high- and
low-density environments. There is a clear offset indicating that the
stellar populations of early-type galaxies in low density environments
are on average about 2$\;$Gyr younger in agreement with previous
results
\citep{Traetal00b,Pogetal01b,Kunetal02b,TF02,CRC03,Proetal04}. This
relative age difference caused by environmental effects is in very
good agreement with the predictions from semi-analytic models of
galaxy formation \citep{KC98a}.

\subsection{Low-mass galaxies - triangles}
The objects of the low-mass tail of the velocity dispersion
distribution (Fig.~\ref{fig:histsig}) with stellar masses
$M_*<10^{10}\Msun$ (Eqn.~\ref{eqn:masssigma}) exhibit all relatively
high Balmer indices, indicating light-averaged ages between 2 and
3$\;$Gyr (triangles in Fig.~\ref{fig:mixgrid}). We recover these
objects with 2-component composite stellar population models based on
the the 'old population' (circles in Fig.~\ref{fig:mixgrid}) the
properties of which are specified in
Eqn.~\ref{eqn:relations}. Extrapolation to the appropriate $\sigma$
yields ages between 6 and 8$\;$Gyr for the base old population.  The
average ages of the low-mass objects in both environments are then
matched by contaminating this base old population with a 10 per cent
contribution (in mass) of a young ($\sim 1.2\;$Gyr) population (see
filled triangles in Fig.~\ref{fig:mixgrid}). As mentioned earlier,
this is certainly not a unique solution, as the contribution in mass
and the look-back time of the star formation event are highly
degenerate and therefore only poorly constrained. Alternatively, also
the addition of a major young component contributing 70 per cent in
mass with an age of 2$\;$Gyr reproduces the observed line indices. We
can conclude, however, that relatively recent star formation in
low-mass early-type galaxies is required to explain the observational
data \citep[see also][]{Traetal00b}.

The model with the minor young component gets some support from the
\aFe\ ratios. The bottom panel of Fig.~\ref{fig:mixgrid} shows that
also the Mg- and Fe- indices are well reproduced by assuming that the
young component is 0.2$\;$dex less \aFe\ enhanced than the base old
population. This clearly favors the option of a young component over
the possible presence of old blue horizontal branch stars as an
explanation for the relatively strong Balmer absorption observed in
low-mass galaxies.  The metal indices of the objects are relatively
weak, we needed to assume the young component to be about $0.2\;$dex
more metal-rich than the base old population.

\subsection{Young intermediate-mass galaxies - squares}
This subpopulation of objects with velocity dispersions between $110$
and $230\;${\rm km/s}, hence $7\cdot 10^9\la M_*/\Msun\la 2\cdot
10^{11}$ (see Eqn.~\ref{eqn:masssigma}), represents a minor fraction
of the whole sample, namely $\sim 14\;$ per cent containing both
elliptical and lenticular types.  There is no obvious dependence on
environmental density and galaxy type. We consider the slight bias
toward more lenticulars in high-density environments (6 S0s out of 8
early-types in high density vs.\ 4 S0s out of 9 early-types in low
density) as statistically not significant given the relatively small
sample size.

The striking feature of these objects is their relatively strong
Balmer line index, yielding light-average ages between 2 and
$4\;$Gyr. To reproduce their \Hb\ indices, we model these galaxies
with a 2-component model adding a minor young component as described
in the previous paragraph.  Different from low-mass galaxies, however,
are the metallicities of the young subcomponent. To match the
relatively strong \MgFe\ index observed, extraordinarily high
metallicities for the young component are required. The smaller the
contribution (in mass) of the subcomponent is, the higher its
metallicity needs to be. The avoidance of unplausibly high
metallicities ($[\ZH]\ga 0.8\;$dex) sets therefore a lower limit to
the weight of the subpopulation. Metallicities below that limit are
ensured, if the young subpopulation contributes at least 20 per cent
in mass. For this lower limit, an age of $1.4\;$Gyr and metallicities
around $[\ZH]\sim 0.8\;$dex are required to reproduce the
observational data. An upper limit can not be constrained.  The high
metallicity of the young component points toward a much higher
effective yield during the formation of the young component in these
intermediate-mass objects, compared to the low-mass objects discussed
above.  This means that chemical evolution and the build-up of metals
is less efficient in low-mass objects, in agreement with the general
picture that objects with shallower potential wells develop galactic
winds easier \citep[e.g.][]{AY87,Ma94,Ed90,ChiCar02}.

Also different from the case of the low-mass galaxies, we had to
assume the young population to have the same \aFe\ ratio like the base
old population, in order to reproduce their Mg- and Fe-indices (bottom
panel of Fig.~\ref{fig:mixgrid}, compare to dotted model lines for
$t=2\;$Gyr). This spoils the interpretation of the enhanced Balmer
line strengths in terms of the presence of an intermediate-age stellar
population. The separation in age between the old and young components
amounts to several billion years, plenty of time for Type~Ia
supernovae to enrich the interstellar medium with iron. Forming out of
this Fe-enhanced material, it will be impossible for the young
component to reach significantly super-solar \aFe\ ratios. Even the
self-enrichment of the young component from Type~II supernovae during
the burst is not sufficient to raise the average \aFe\ ratio of the
newly born stellar populations independent of the burst timescale as
shown in \citet{TGB99}.

The interpretation of the observed strong Balmer line indices in terms
of recent star formation could be recovered under the rather contrived
assumption that the Type~Ia supernova ejecta originating from stars of
the base old population have been lost selectively in galactic winds
before the formation of the young subpopulation.  Alternatively, the
latter must have formed with a stellar initial mass function (IMF)
significantly flatter than Salpeter. A flat IMF slope would naturally
reproduce the very high metallicity of the young component derived
here. On the other hand, compelling evidences for systematic
variations of the IMF slope in starbursts have not been found so far
\citep{Lei98}.

But we have to be cautious here.  Relatively high \aFe\ ratios are
needed for the young subpopulation, because in the models \Mgb\
decreases significantly with decreasing age at highest metallicity,
and it is possible that the fitting functions are in error in this
regime owing to bad input stellar data. On top of this, the
extraordinarily high metallicities of the young subcomponent
($[\ZH]\sim 0.8$) require extrapolation well above the highest
metallicity model ($[\ZH]=0.67$), which makes the model indices very
unreliable. The relatively low \Mgb\ index of the model, which causes
the need for the relatively high \aFe\ ratio, might be an artifact of
bad calibration. To conclude, lower values for the young subpopulation
cannot be excluded, which rehabilitates the possibility of the
presence of an intermediate-age population.

Alternatively, the strong Balmer lines might be the result of blue
horizontal branch stars rather than a young population, which we shall
discuss in detail below.

\subsection{Young massive galaxies - diamonds}
Finally, the diamonds indicate a subclass of very massive early-type
galaxies ($M_*>10^{11}\;\Msun$) that have strong \Hb\ and strong
\MgFe\ absorption indices, implying light-averaged ages between 2 and
$5\;$Gyr, and very high metallicities of $[\ZH]\ga 0.67\;$dex.  From
Fig.~\ref{fig:mixgrid} it can be seen that the occurrence of these
objects is restricted to high-density environments. Moreover, they are
only lenticular galaxies, no ellipticals with such low average ages
and high metallicities are found (see Fig.~\ref{fig:mixgrid}).  The
vast majority (80 per cent) of all S0 galaxies with $\sigma>200\;$km/s
in our sample and the two Coma cD galaxies belong to this object class
(see also \citealt{Kun00,Pogetal01b,Mehetal03}).

The metallicities are more than $0.2\;$dex above the average
metallicity of the most massive objects in the old base population
constrained by Eqn.~\ref{eqn:relations}. Such a significant increase
in average metallicity can obviously not be obtained in a 2-component
model with only 10--20 per cent of a metal-rich subpopulation, unless
unreasonable high metallicities above ten times solar are assumed. We
therefore have to increase the weight of the young population. The
filled diamonds shown in Fig.~\ref{fig:mixgrid} are models in which
the old base population is 'perturbed' with a $2.5\;$Gyr young
population with metallicities as high as $[\ZH]\sim 0.7\;$dex
contributing 70 per cent in mass. Lower metallicities can be adopted
only under the assumption that an even larger fraction of the stellar
populations belongs to the second component.

Interestingly, like in the previous case these objects have very high
\aFe\ ratios. For the young component we need to adopt \aFe\ ratios as
they are given by Eqn.~\ref{eqn:relations}, hence $[\aFe]>0.2$ for
$\sigma>200\;$km/s. This fits well into the picture that the young
average ages are not the result of minor recent star formation, but
rather the consequence of a recent very massive star formation event
at a redshift around $z=0.3$ forming the majority of the galaxy's
stellar populations on a short timescale.  This would imply that
massive S0s and cD galaxies in clusters form very recently, while
ellipticals of comparable mass contain only old stellar populations.
This explanation seems contrived, and it remains to be seen if it can
be understood in the framework of galaxy formation models.

\subsection{Blue horizontal branch stars at high metallicities}
Alternatively, the strong \Hb\ indices discussed above might be caused
by the presence of metal-rich, blue horizontal branch stars (see
Introduction). Such blue horizontal branch stars may originate from
either a metal-poor subpopulation as discussed in \citet{MT00} or
metal-rich populations with blue horizontal branch morphologies due to
enhanced mass loss along the red giant branch evolutionary phase. Here
we consider this latter option.

\begin{figure*}
\begin{center}
\includegraphics[width=0.8\textwidth]{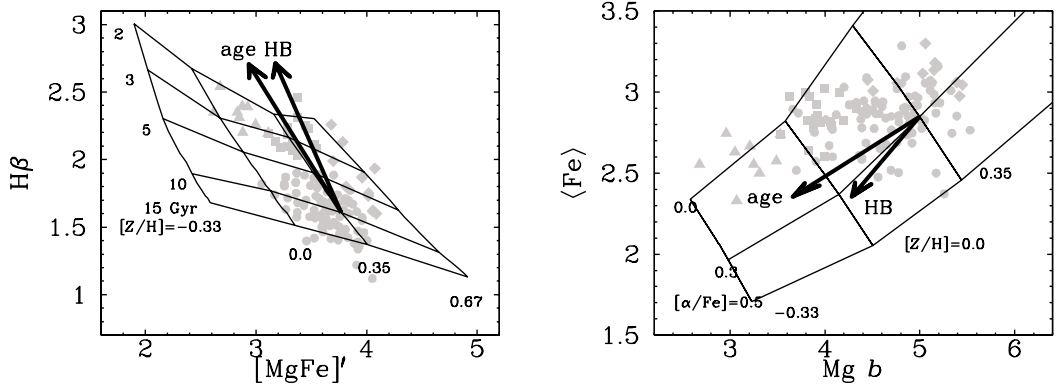}
\end{center}
\caption{Comparison of the effects of age and blue horizontal branch
on the absorption indices \Hb,\Mgb, and \Fe. Symbols and model grid
like in Fig.~\ref{fig:mixgrid}. The arrows labeled 'HB' show the index
changes of a $10\;$Gyr population ($[\ZH]=0.35,\ [\aFe]=0.3$, RGB mass
loss parameter $\eta=0.33$) when a blue horizontal branch $\eta=0.55$
is considered. For comparison, the effect of reducing the age to
$1.5\;$Gyr is indicated by the arrows labeled 'age'.}
\label{fig:mgfeHB}
\end{figure*}

\subsubsection{The origin of blue horizontal branches}
The mass-loss along the red giant branch can be theoretically
understood as stellar winds being produced by the acoustic flux
generated by the superadiabatic convection in the outer convective
zone of the red giant \citep{FR75}. The fraction of acoustic energy
which is used to remove the stellar envelope is not known from
physically motivated, first principles. An alternative is the
empirical determination of the mass loss rate, which has been done by
\citet{Reimers75} studying the behavior of the circumstellar calcium K
line in a large number of red giants and supergiants. The
normalization of the resulting formula, however, suffers from very
large uncertainties.

\citet{FR76} combined the two approaches and introduced an efficiency
factor $\eta$ for Reimer's formula. They calibrated the mass loss rate
with observational constraints like the horizontal branch morphology
\citep{Rood73}, the limiting initial mass for carbon ignition ($\sim
4$~\Msun\ for the Hyades \citealt{Heuvel75}), and finally the solar
mass loss rate. An efficiency $\eta\approx 0.33$ turned out to be the
most appropriate choice. This value is adopted in the models of
\citet{Ma98,Ma05}. Note that the observational uncertainties in
Reimer's formula allow this efficiency factor to lie in the range
$0.3\la\eta\la 3$. The problem is that the calibration has been
carried out at sub-solar metallicity, and deviations from this
canonical value, in particular at higher metallicities, cannot be
excluded \citep{GR90}.

\subsubsection{Effect on the line indices}
The models of TMB are based on \citet{Ma98,Ma05}, thus rely on the
calibrated value $\eta=0.33$. Here we additionally compute models with
the slightly larger efficiency factor $\eta=0.55$, which is well
within the range of uncertainty allowed by the Reimer's formula. These
models have bluer horizontal branch morphologies, and therefore
predict significantly stronger Balmer indices, and slightly weaker
metallic indices. This is shown in the left panel of
Fig.~\ref{fig:mgfeHB} by the arrow labeled 'HB' for a $10\;$Gyr old
population. The arrow labeled 'age' shows the location of a model with
age $t=1.5\;$Gyr, which yields the same \Hb\ value, for comparison.
The steeper slope of the 'HB' model illustrates the weaker sensitivity
of the metallic line indices to HB morphology than to age. The
relative effect of \Mgb\ and \Fe\ is shown in the right panel of
Fig.~\ref{fig:mgfeHB}. Relative to \Fe\, the \Mgb\ index responds
stronger to a decrease of age than to a blueing of the horizontal
branch. As a consequence, the slope of the 'HB' model is steeper also
in the \Mgb-\Fe\ diagram. Note that the slope is comparable to the
model varying metallicity at a constant \aFe\ ratio and age.

\subsubsection{2-component model}
These results will mainly affect the age determination. To test this,
we have constructed a 2-component model in which the base old
population is contaminated with 50 per cent of an old metal-rich
population with blue horizontal branch stars.  The location of both
the squares and the diamonds in Fig.~\ref{fig:mixgrid} is reached by
these models, hence without invoking the presence of intermediate-age
populations. We had to assume the metallicity of the second component
to be about $0.2$--$0.4\;$dex higher than the average of the base
population, while the \aFe\ ratio needs to be about $0.1\;$dex
lower. Hence, the blue HB option requires the assumption that
preferentially the metal-rich stars in a composite stellar population
develop blue horizontal branches. The slightly lower \aFe\ ratios
would fit into this picture, as the most metal-rich populations must
have formed latest and should therefore be least \aFe\ enhanced.

\subsubsection{Young vs.\ blue horizontal branch}
To conclude, none of the two options---intermediate-age population or
blue horizontal branch---can be excluded, further observational
constraints to distinguish the two alternative explanations are
required. 

If intermediate-age populations are present in these galaxies, they
should be detectable by their very red near-infrared colors, because
of the major contribution to the bolometric from asymptotic giant
branch stars at ages between $0.5$ and $2\;$Gyr
\citep{Ma98,Maretal01,Maraston04,Ma05}.

Un-ambiguous indicators of blue horizontal branches, instead, still
need to be found.  The Ca{\sc II} index defined by \citet{Rose84},
even though promising, does not solve the problem as it is still
degenerate between gravity and metallicity effects.  \citet{Schetal04}
show that in metal-poor globular clusters the higher-order Balmer line
index \HdF\ is more sensitive to the presence of blue horizontal
branches than \Hb. Hence, the index ratio \HdF/\Hb\ may be a tool to
break the degeneracy between age and horizontal branch morphology. It
is not clear, however, if this effect is still present at solar
metallicities and above, which are relevant to massive
galaxies. Moreover, in \citet{TMK04} we show that at such high
metallicities the \HdF/\Hb\ index ratio becomes very sensitive to the
\aFe\ ratio in the sense that \HdF/\Hb\ increases with \aFe. If blue
horizontal branches ought to be detected in early-type galaxies by
means of the \HdF/\Hb\ ratio, the dependence on \aFe\ certainly has to
be taken into account. It will be interesting to make these checks for
the galaxy sample discussed here. Such an analysis will be subject of
a future study.

\section{The origin of the Mg-$\sigma$ relation}
\label{sec:mgsigma}
The correlations of the stellar population parameters age, total
metallicity, and \aFe\ ratio with central velocity dispersion shown in
Fig.~\ref{fig:popsall} and Eqn.~\ref{eqn:relations} allow us to
constrain the origin of the well-known Mg-$\sigma$ relation (top
panels of Fig.~\ref{fig:indexall}). We recall that the correlations
found here suggest that all three parameters correlate with $\sigma$,
hence support the Mg-$\sigma$ relationship.

\begin{figure}
\begin{center}
\includegraphics[width=0.9\linewidth]{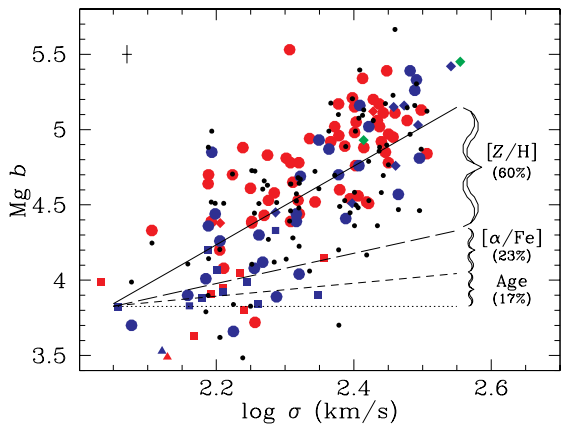}
\caption{\Mgb-$\sigma$ relation of the whole sample high density and
low density environments combined. Colored symbols are observational
data (red: elliptical, blue: S0, green: cD), median 1--$\sigma$ error
bars are shown. Black symbols Monte Carlo simulations (symbol types
like in Fig.~\ref{fig:mixgrid}).  The solid line is the \Mgb-$\sigma$
relation resulting from the relationships given in
Eqn.~\ref{eqn:relations} averaging over high and low density
environments. The short-dashed line shows the resulting \Mgb-$\sigma$
relation if only age is assumed to be the driving parameter adopting
Eqn.~\ref{eqn:relations}. The long-dashed line demonstrates the effect
of additionally adding the dependence on \aFe\ ratio from
Eqn.~\ref{eqn:relations}. }
\label{fig:mgsig}
\end{center}
\end{figure}
Fig.~\ref{fig:mgsig} shows the observed \Mgb-$\sigma$ relation of the
whole sample high density and low density environments combined. It
can be verified from Fig.~\ref{fig:indexall} that the environment has
a negligible effect on the Mg-$\sigma$ relation in agreement with
previous findings based on much larger samples
\citep{Beretal98,Coletal99,WC03}. Circles are the 'old population'
representing the major fraction of the sample, triangles, squares, and
diamonds are the 'young outliers' discussed in the previous section
(see also Fig.~\ref{fig:mixgrid}). Red and blue symbols are elliptical
and lenticular galaxies. Green symbols are cD galaxies. The black
symbols are the Monte Carlo realizations.  The solid line is the
\Mgb-$\sigma$ relation that results from the relationships given in
Eqn.~\ref{eqn:relations} (averaging over both environments). The
short-dashed line shows the resulting \Mgb-$\sigma$ relation if only
age is assumed to be the driving parameter adopting
Eqn.~\ref{eqn:relations}. The dependencies of both total metallicity
and \aFe\ ratio with $\sigma$ are suppressed. The long-dashed line
demonstrates the effect of additionally adding the dependence on \aFe\
ratio from Eqn.~\ref{eqn:relations}.

It can be seen that total metallicity has by far the largest share (60
per cent) in shaping the relation, followed by the \aFe\ ratio
contributing only 23 per cent to the observed slope. The contribution
from age is about 17 per cent, hence is smallest but comparable to the
one of \aFe.  We emphasize that these numbers do not vary
significantly with environmental density. As can be inferred from
Figs.~\ref{fig:indexall} and~\ref{fig:popsall}, the scatter of the
\Mgb-$\sigma$ relation among the old population (circles) results from
intrinsic scatter in all three stellar population parameters.  The
\Fe-$\sigma$ relation is much shallower (see Fig.~\ref{fig:indexall})
because the increasing \aFe\ ratio with $\sigma$ diminishes the
\Fe-index, compensating largely the increase due to increasing
metallicity (see also \citealt{Mehetal03}).

This result fits very well to the fact that also the color-magnitude
relation appears to be driven by metallicity rather than age
\citep{KA97,SED98,Teretal99,Vazetal01,Meretal03}, even though the
effect of the \aFe\ ratio variations on colors is still not very well
understood \citep{TM03}.

Finally, we briefly examine the location of the 'young outliers' of
Fig.~\ref{fig:mixgrid} (triangles, squares and diamonds).  The
low-mass and intermediate-mass objects with young light-averaged ages
(triangles and squares) are clearly off the relation and exhibit
systematically lower index values (by $\sim 0.5\;$\AA) at a given
$\sigma$. Their slightly larger light-averaged metallicities (see
Fig.~\ref{fig:mixgrid}) do obviously not suffice to bring them back on
the relationship. Different the class of massive, apparently young
Coma cluster lenticulars (diamonds). Their extremely high
light-averaged metallicities compensate totally for the decrease in
\Mgb\ caused by the young light-averaged ages. As far as the
\Mgb-$\sigma$ relation is concerned, these objects are
indistinguishable from the main body of the sample.

\section{Formation epochs}
\label{sec:sfh}
\begin{figure*}
\centerline{\includegraphics[width=0.6\linewidth]{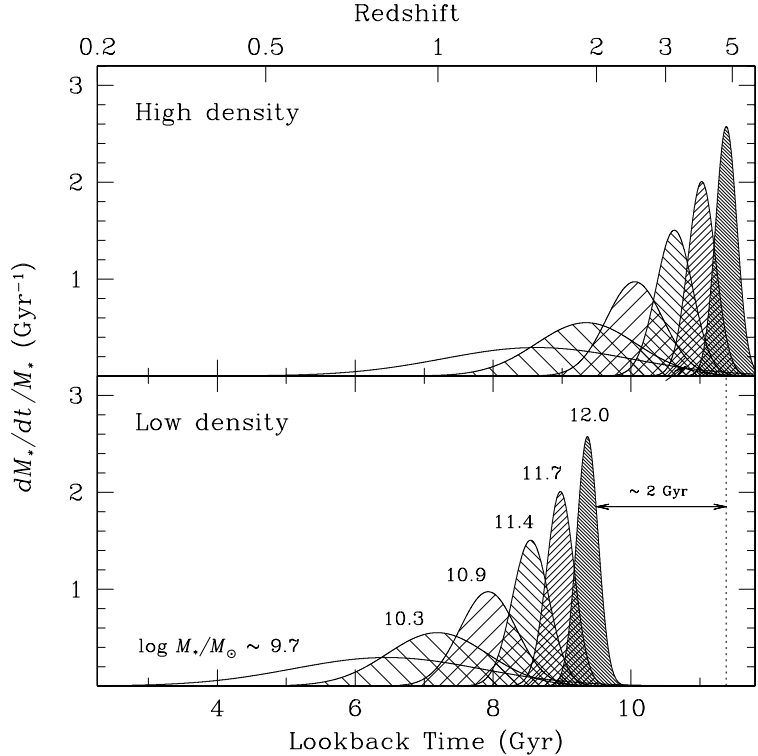}}
\caption{Star formation histories of early-type galaxies as a function
of their stellar masses $M_*$ (see labels). The stellar masses $M_*$
correspond to the velocity dispersions $\sigma=100,\ 140,\ 190,\ 240,\
280,\ 320\;$km/s (see Eqn~\ref{eqn:masssigma}). The star formation
histories are derived from the mean ages and [\aFe] ratios shown in
Fig.~\ref{fig:popsall} and Eqns.~\ref{eqn:relations} (or
\ref{eqn:massrelations}) and~\ref{eqn:sfh}.  Redshifts assume
$\Omega_m=0.3,\ \Omega_{\Lambda}=0.7,\ H_0=75$ km/s/Mpc
\citep{Speetal03}. The dotted lines marks the average age of a high
density objects with $M_*=10^{11}\;\Msun$ ($\sigma=320\;$km/s) for
comparison.}
\label{fig:sfh}
\end{figure*}
The relations shown in Fig.~\ref{fig:popsall} (see also
Eqns.~\ref{eqn:relations} and~\ref{eqn:massrelations}) can be used to
constrain the epochs of the main star formation episodes and the star
formation timescales for early-type galaxies as a function of their
mass and environmental densities. It should be emphasized that the
majority of the objects of the present sample obeys these
relationships. Only 15 per cent of the sample (squares in
Fig.~\ref{fig:mixgrid}) requires a 20 per cent contribution from
intermediate age populations on top of the base old population
described by Eqn.~\ref{eqn:relations}, and further 9 per cent
(diamonds) are best modeled assuming 70 per cent young
populations. The young low-mass galaxies (triangles in
Fig.~\ref{fig:mixgrid}) do not play a significant role in the total
mass budget of early-type galaxies, the latter being dominated by
$L^*$ galaxies with $M_*\sim 10^{11}\;\Msun$ situated at the turnover
of the luminosity (mass) function \citep{Beletal03a,Baletal04}.  This
implies that about 90 per cent of the total stellar mass hosted by
early-type galaxies is well represented by the relationships given in
Eqn.~\ref{eqn:relations}.

\subsection{Relating timescale and \aFe\ ratio}
The key point of this analysis is that the \aFe\ ratio lifts the
degeneracy between formation epoch and formation timescale. This is
done by relating the observationally derived luminosity-averaged \aFe\
ratios of the stellar populations with timescales of their formation.

For this purpose we simulate the formation and chemical evolution of a
set of toy galaxies assuming various star formation histories
characterized by different timescales similar to the exercise
presented in \citet{TGB99}. 

We start with a gas cloud of primordial chemical composition. By
imposing a star formation rate as a function of time, which is given
by the specific star formation history assumed, we let the gas cloud
form stars. These stars enrich the gas of the cloud (the interstellar
medium) with heavy elements produced mainly in supernova
explosions. For the aim of this paper we follow the chemical
enrichment of $\alpha$ and Fe-peak elements. A key point is that we
take the delayed enrichment of Fe from Type~Ia supernovae into account
using the prescription of \citet[][see \citealt{TGB98} for more
details]{GR83}. Stellar populations successively form out of more and
more chemically enriched gas, resulting in the build-up of a composite
stellar population. Every generation of stars in this composite will
have the chemical composition, hence \aFe\ ratio, of the gas out of
which it had formed.  We examine this population $12\;$Gyr after the
formation of the first stellar generation, and calculate its $V$-light
averaged [\aFe] ratio, which corresponds to the quantity
observationally derived in this study. The weight in $V$-luminosity of
the single stellar populations as a function of their ages is adopted
from the $M/L$ ratios of \citet{Maraston04}. The IMF slope is kept
fixed to Salpeter. Like in \citet{TGB99}, the simulations are carried
out assuming a closed box, which implies that selective mass-loss is
assumed not to occur and thus not to affect the \aFe\ ratio.

The star formation histories in the different runs are chosen to be
Gaussians of various widths $\Delta t$ (FWHM).  We find the following
linear relationship between \aFe\ ratio and $\log\Delta t$:
\begin{equation}
[\aFe]\approx 1/5 - 1/6\ \log\ \Delta t\ .
\label{eqn:sfh}
\end{equation}
The equation reflects what has already been shown in \citet{TGB99},
namely that a ratio $[\aFe]=0.2$ of a composite stellar population
requires formation timescales $\Delta t\la 1\;$Gyr.  The larger
$\Delta t$, the lower is the final \aFe\ ratio of the stellar
population because of the late enrichment of Fe from Type~Ia
supernovae.  Star formation extended over the Hubble time yields an
\aFe\ close to solar. We note that this link between \aFe\ ratio and
star formation timescale is in very good agreement with the more
detailed chemical evolution simulations of \citet{PM04}.

The combination of Eqns.~\ref{eqn:relations}
and~\ref{eqn:massrelations} with Eqn.~\ref{eqn:sfh} then yields
correlations of $\Delta t$ with $\sigma$ and mass for early-type
galaxies valid in both high and low environmental densities.
\begin{eqnarray}
\log\Delta t &\approx& 3.44 - 1.68\ \log\sigma\ \\\nonumber
             &\approx& 3.67 - 0.37\ \log M_*/\Msun
\label{eqn:timescale}
\end{eqnarray}
Eqns.~\ref{eqn:relations} and~\ref{eqn:massrelations} further give the
light-average ages as a function of $\sigma$, constraining the epoch
at which the object has formed.

\subsection{The star formation histories}
The resulting star formation histories for various galaxy masses and
velocity dispersions are shown in Fig.~\ref{fig:sfh}. The x-axis is
lookback time, the upper x-axis gives the corresponding redshifts
assuming $\Omega_m=0.3,\ \Omega_{\Lambda}=0.7$, and $H_0=75$~km/s/Mpc
\citep{Speetal03}. The top and bottom panels show the results for
dense and loose environments, respectively.

It should be kept in mind that Fig.~\ref{fig:sfh} is based on average
ages and \aFe\ ratios at a given mass. As discussed in
Section~\ref{sec:results}, the data require a scatter in \aFe\ ratio
of the order $0.05-0.07\;$dex (see Fig.~\ref{fig:residuals}), which
translates into a spread in star formation timescale of about a factor
2. The observed spread in average age is consistent with the a scatter
of about 20--25 per cent (see Fig.~\ref{fig:residuals}). For a cluster
early-type galaxy with stellar mass $M_*\sim 5\cdot 10^{11}\;\Msun$
($\sigma=280\;$km/s), for instance, this implies a spread about the
typical formation redshift at $z\sim 3$ between $z\sim 2$ and $z\sim
5$. The blurring caused by these effects should be kept in mind for
the interpretation of the diagrams in Fig.~\ref{fig:sfh}, which are
meant to be a schematic illustrations of the mean formation epochs.

\subsection{Dependence on galaxy mass}
The plots summarize what has been already discussed in the previous
sections. The more massive the galaxy is, the faster and earlier form
its stellar populations. Indeed \citet{PM04} show that the observed
relation between \aFe\ and $\sigma$ can be best reproduced by a model
in which gas accretion and star formation timescales decrease with
increasing galaxy mass.  Only low-mass galaxies exhibit significant
star formation at redshifts below $z\la 1$ as also concluded by
\citet{BE00} and more recently \citet{Deletal04}.  These results are
also in very good agreement with the star formation histories found by
\citet{Gavetal02} based on spectrophotometric data. Interestingly, the
link between star formation histories and galaxy mass shown in
Fig.~\ref{fig:sfh} agrees well with the predictions from
N-body-tree-SPH simulations by \citet{ChiCar02}, which are based on a
monolithic collapse inside a dark matter halo rather than hierarchical
structure formation.

Note that we have assumed continuous star formation in order to keep
the number of free parameters low. The principal aim of this exercise
is to illustrate the typical timescale of the star formation histories
in galaxies as a function of their masses. In reality, star formation
in individual objects will be less smooth and will be shaped by
several bursty star formation episodes \citep[see][]{ChiCar02}. The
functions of Fig.~\ref{fig:sfh} thus have to be understood as
integrals at a given velocity dispersion over bursty star formation
histories. In particular the extended star formation histories of the
low-mass objects might rather be understood in terms of episodically
returning star forming bursts, similar to what is found in dwarf
galaxies of the Local Group \citep{Mateo98}.

In massive systems, the burst might well be triggered by merger
events.  Indeed, a considerable fraction of massive ellipticals have
kinematically decoupled cores \citep{Be88,FI88}, the presence of which
requires the merger to be accompanied by dissipational processes and
star formation. Interestingly, for a number of ellipticals with
kinematically decoupled cores, it has been found that their stellar
populations are homogeneously old and \aFe\ enhanced
\citep{SB95,Mehetal98,Davetal01}, indicating the merger to have
happened at early times, in agreement with the formation histories of
Fig.~\ref{fig:sfh} (a counter-example is NGC~2865, \citealt*{HCB99}).

\subsection{Influence of the environment}
The second main result is that galaxy formation appears delayed by
$\sim 2\;$Gyr in low density environments confirming previous findings
\citep{Traetal00b,Pogetal01b,Kunetal02b,TF02,CRC03,Proetal04}.

This is in agreement with studies of the field galaxy population at
intermediate redshifts around $z\sim 1$
\citep{vDoketal01,Treetal02,Gebetal03}. While the formation of massive
early-type galaxies in clusters sets in at redshifts around $z\sim 5$
and is completed (in terms of their stellar populations) at about
$z\sim 2$, a considerable fraction of massive galaxy formation in low
density environments occurs at lower redshifts between $z\sim 1$ and
$z\sim 2$.  We recall here that the \aFe-$\sigma$ relation does not
depend on environmental density. The timescale of star formation in an
object is determined by its (final) mass rather than the environment.
As a consequence, massive early-type galaxies in low density
environments form later but on similar timescales like their
counterparts in high environmental densities as illustrated in
Fig.~\ref{fig:sfh}.

\section{Discussion}
\label{sec:discussion}

\subsection{The local abundance pattern at high metallicities}
In a recent study, \citet{Proetal04} argue that the correlation of
\aFe\ with velocity dispersion is an 'artefact of incomplete
calibration (of the models) to the Lick system'. The basis for their
conclusion is the recent finding of \citet{BFL04}, that the Milky Way
disk stars with super-solar metallicities do not have solar but
sub-solar O/Fe ratios, O/Fe decreasing with increasing Fe/H \citep[see
also][]{PBG03,Prietal04}. Stellar population models, being calibrated
with these stars, are then likely to reflect sub-solar O/Fe ratios at
high metallicities.  If the element Oxygen is taken as representative
for all $\alpha$-elements, this would obviously lead to an
overestimation of the O/Fe ratio (or \aFe\ ratio) in stellar
populations with super-solar metallicities like early-type
galaxies. \citet{Proetal04} show that the correction for this effect
makes the \aFe-$\sigma$ relation disappear.

However, there is a further very important (and somewhat puzzling)
point shown in \citet{BFL04}. All the other $\alpha$-elements, Mg, Si,
Al, Na, Ca, and Ti, do not show this pattern, but level-off at a solar
X/Fe ratio in stars with super-solar metallicities \citep{Fu98,BFL03}.
Only the element oxygen (and maybe also carbon, \citealt*{TEG04})
sticks out of this homogeneous behavior.  This differential pattern
between Oxygen and the other $\alpha$-elements cannot be neglected in
models taking the local abundance pattern at high metallicities into
account. This is particularly important in the context of this work,
as the \aFe\ ratios presented here are derived from the absorption
line indices \Mgb\ and \Fe, which are sensitive mostly to the elements
Fe and Mg, but {\em not} O \citep{TB95,KMT05}.

We investigated this issue constructing two additional flavors of
stellar population models that account for this local abundance
pattern at high metallicities. The first model takes oxygen as
representative for all $\alpha$-elements (like in
\citealt{Proetal04}), in the second model oxygen is detached from the
rest of the $\alpha$-elements as discussed above.  The resulting
models in the index-index planes and the stellar population parameters
derived with these models are shown in Figs.~\ref{fig:biasgrid1}
and~\ref{fig:biasgrid2} (equivalent to Fig.~\ref{fig:datagrid}) and
Figs.~\ref{fig:popsall2} and~\ref{fig:popsall3} (equivalent to
Fig.~\ref{fig:datagrid}) in the Appendix. The results are discussed in
detail there.

\smallskip
To summarize, we reproduce the result of \citet{Proetal04} with the
model in which oxygen and $\alpha$-element are not separated. The
\aFe-$\sigma$ relation disappears, the \ZH-$\sigma$ relation becomes
tighter, while the ages remain unchanged.  When oxygen is detached
from the other $\alpha$-elements as suggested by observations, the
picture is entirely different as explained in detail in
Section~\ref{sec:bias}. The correction for the local abundance pattern
has no significant impact on the model, when the non-solar O/Mg ratios
of metal-rich stars in the solar neighborhood are taken into account.
The model basically moves back to its original position.  As a
consequence, the stellar population parameters derived with this model
are inexcellent agreement with the ones presented in this work (see
Fig.~\ref{fig:popsall3}).

To conclude, the results of this paper are robust against the
correction for the local abundance pattern at high metallicities. In
particular, the \aFe-$\sigma$ relation found here is real and is not
caused by calibration artefacts.  What remains to be understood is the
discrepancy between oxygen and the other $\alpha$-elements in the
metal-rich stars of the solar neighborhood.

As discussed in \citet{BFL04}, a possible explanation could be that
Type~Ia supernovae do produce a significant amount of magnesium, which
keeps the Mg/Fe ratio solar, while O/Fe decreases with increasing
metallicity. This would also fit to the slightly higher \OFe\ than
Mg/Fe ratios in metal-poor stars of the Milky Way \citep{BFL04}.  In
this case, the standard model for Type~Ia nucleosynthesis \citep*[the
so-called 'W7' model;][]{NTY84} needs to be revised. This leads to the
speculation that the importance of Type~Ia enrichment may increase
steadily with atomic number of the element up to Fe and Ni. If this is
the case, the higher the atomic number of the element, the flatter
should be the X/Fe-Fe/H relation in the Milky Way.  In particular
elements with high atomic number should have lower X/Fe ratios in the
metal-poor halo stars, which has not been confirmed in the review of
\citet{McW97}. Maybe new data sets will revolutionize this
statement. If true, the low Ca/Mg ratios of early-type galaxies
\citep[e.g.][]{TMB03b} would naturally be explained in terms of
delayed Type~Ia supernova enrichment \citep{PM04}.

The relevant inference for this paper would be that the formation
timescales discussed in Section~\ref{sec:sfh} (Eqn.~\ref{eqn:sfh}) are
underestimated. A detailed investigation and quantification of this
effect goes far beyond the scope of this paper, and in particular
cannot be performed in a straightforward way as the appropriate
Type~Ia nucleosynthesis calculations are not available. 
Note however,
that the high \aFe\ ratios derived here for massive galaxies,
suggesting short star formation timescales, are coupled with old
average ages, which further reinforces this interpretation.  The
results of this work suggest star formation timescales as short as
only a few $10^{8}$ years in the most massive early-type galaxies with
$M_*\sim 10^{12}\;\Msun$. It would certainly be desirable if such
extreme timescales were slightly relaxed by the additional
contribution from Type~Ia supernovae to the enrichment of the heavier
$\alpha$-elements.  It should be emphasized that a fundamental
conclusion of this paper is that low-mass galaxies have more extended
formation timescales. This purely differential result is not affected,
if Type~Ia supernovae represent an additional source of
$\alpha$-elements.

\subsection{Comparison to early results}
The findings discussed in this paper are supported by a number of
evidences from the tightness and the redshift evolution of the scaling
relations of early-type galaxies \citep[e.g.][see review by
\citealt{Peebles02}]{BLE92,Araetal93,BBF93,RC93,BZB96,KBB99,Zieetal99,Sagetal00,Zieetal01}.
It is known since these early (and some more recent) works that the
formation ages of the stellar populations in elliptical galaxies have
to be above redshifts $z\ga 2$. New in this paper is the detailed
disentanglement of star formation histories as a function of galaxy
mass and the distinction between dense and loose environments.

We confirm such high formation redshifts for intermediate-mass and
massive ($M_*\ga 10^{10}\;\Msun$ or $\sigma\ga 140\;$km/s) early-type
galaxies in dense environments in good agreement also with the
recently discovered relationship between globular cluster color and
galaxy luminosity \citep*{SBF04} and the $K$--$z$ relation of powerful
radio galaxies \citep{Rocetal04}.  

We find, however, that ellipticals in looser environments form at
somewhat lower redshifts between $z\sim 2$ and $z\sim 1$. Low-mass
galaxies ($M_*\la 2\cdot 10^{10}\;\Msun$ or $\sigma\la 140\;$km/s) in
low density regions apparently suffer from significant star formation
episodes below redshift $z\sim 1$.  This paper further adds the
conclusion that the stellar populations of more massive early-type
galaxies form earlier and faster. This result confirms early, more
qualitative estimates by \citet{BZB96} based on the redshift evolution
of the Mg-$\sigma$ relation, who had already noted that the formation
redshifts of the most massive early-type galaxies in clusters may be
even above $z\ga 4$. Here we find that while the formation epoch is
delayed in low-density environments by about 2$\;$Gyr, the timescale
is independent of the environmental density and depends only on the
galaxy's mass concentration.

\subsection{The stellar mass density at $z>1$}
The above evidences provide consistent constraints on the epochs when
the stellar populations of early-type galaxies have formed. Our
results imply that at redshift around $z\sim 1$, the formation of the
total stellar mass in early-type galaxies with $M_*\ga 2\cdot
10^{10}\;\Msun$ or $\sigma\ga 140\;$km/s in both high and low density
environments is almost completed. Stellar mass growth in early-type
galaxies below redshift $z\sim 1$ is restricted to low-mass objects in
low density environments. As the latter make up only a small fraction
of the total mass of a galaxy population, despite them being numerous,
we can infer that stellar mass growth in early-type galaxies is
largely completed around redshift $z\sim 1$.

This galaxy type is expected to host between 1/2 and 3/4 of the total
stellar mass in the universe \citep{Renzini99,Beletal03a}. As also
massive disks are already present at redshifts between 1 and 2
\citep{Ravetal04,SCM04}, this implies that at least 50 per cent of the
total stellar mass density must have already been formed by $z\sim 1$.
Our results further suggest that a significant increase of the stellar
mass density between redshifts 1 and 2 should be present caused mainly
by the field galaxy population, while at redshifts around 3 the
stellar mass density should be dominated by galaxies in high density
environments. Depending on the exact partition between objects in high
and low environmental densities, this implies the stellar mass density
around redshift 3 to be as low as 10--30 per cent of the local value.
Observational estimates of the total stellar mass density as a
function of redshift are in overall good agreement with this
result. Recent studies are consistent with only mild evolution of the
order $20-30$ per cent up to $z\sim 1$ followed by a significant drop
in stellar mass density between redshifts 1 and 2
\citep{BE00,Cohen02,Beletal03a,Dicetal03,Fonetal03,Rudetal03,Fonetal04,Glaetal04}. Our
result suggests that the significant mass growth in the early-type
galaxy population below $z\sim 1$ observed by \citet{Beletal04} is
most likely caused by less massive objects.

\subsection{Climbing up the redshift ladder}
One fundamental consequence from the results of this paper is the
prediction that very massive objects $M_*\ga 10^{11}\;\Msun$) with
very high star formation rates up to several $1000\;$\Msun/yr should
be present between redshifts 2 and~5.

Such objects have indeed been recently found.  At least part of the
SCUBA sources at $2\leq z\leq 3$ turn out to be massive star forming
galaxies with extremely high star formation rates up to several
$1000\;$\Msun/yr \citep{Smaetal02,Genetal03,Tecetal04}. These objects
are likely to be the precursors of the most massive ellipticals
($M_*\approx 10^{12}\;\Msun$) forming in a violent star formation
episode at high redshift. Ly-break galaxies, instead, exhibit more
moderate star formation rates of the order a few $10\;$\Msun/yr
\citep{Petetal01,Baketal04}, and may therefore be the precursors of
less massive early-type galaxies in Fig.~\ref{fig:sfh} (($M_*\approx
10^{11}\;\Msun$) and/or spiral bulges. Moreover, \citet{Petetal02}
find cB58 (a lensed galaxy at redshift $z=2.73$, \citealt{Seietal98})
to have a significantly $\alpha$-enhanced {\em interstellar medium},
which is in good agreement with the \aFe-enhancement derived in this
paper for the {\em stellar population} of local ellipticals.

A further direct inference from Fig.~\ref{fig:sfh} is that massive
old, passively evolved galaxies should exist at intermediate redshifts
between 1 and 2. Again, these objects have been found. Medium-deep
infrared surveys detect a relatively large number of so-called
extremely red objects (EROs) between $z\sim 1$ and 1.5, whose SEDs
best fit evolved stellar populations
\citep{PM00,Dadetal00,Cimetal03,Saretal03,VJ04,Mouetal04,Capetal04,Cimetal04,Saretal04}.
The evolved $K$-selected EROs mentioned above show evidence for strong
clustering \citep{Dadetal03}, hence are likely candidates for being
the precursors of the cluster objects shown in the top panel of
Fig.~\ref{fig:sfh}. In a recent paper, \citet{Pieetal04} show that the
colors of some 'evolved' EROs are also consistent with a dusty {\em
post}-starburst population with ages around $0.5\;$Gyr, which fits to
the star formation histories shown in Fig.~\ref{fig:sfh} for the low
density environment. It would be interesting to investigate the
clustering properties of this subclass of EROs, to verify that they
preferably are found in low density environments.  Also in agreement
with our result is the recent finding in $K$-selected galaxy surveys
that the most luminous objects are best fit by older stellar
populations at all redshifts up to $z\sim 1$, which implies the
typical $M/L$ ratio (in $K$) of more massive galaxies to be larger
than that of less massive ones \citep{Fonetal04,Glaetal04}.

Finally, deeper redshift surveys are now finding evolved objects even
at redshifts around 2 and 3
\citep{Fraetal03,CM04,Mccaretal04,Capetal05} and massive star forming
objects around redshifts 2 and 5 \citep{Totetal01,LB03,deMeletal04} up
to $z\sim 6.5$ \citep{Kuretal04}, again in good agreement with the
prediction from the properties of local galaxies made in this paper.

\smallskip
To conclude, it seems the two approaches, archaeology of stellar
populations in local galaxies and redshift surveys, are converging to
a consistent picture in which massive galaxy formation occurs at
fairly high redshifts accompanied by violent star formation.

\subsection{Models of galaxy formation}
This consistent picture of high formation redshifts of the stellar
populations in massive (early-type) galaxies severely challenges the
conventional semi-analytic models of galaxy formation
\citep{KWG93,SP99,Coletal00,Hatetal03,Menetal03}, and their
modification has become unavoidable. As reviewed by
\citet{Somerville04}, recent renditions of both semi-analytic models
and hydrodynamical simulations of hierarchical galaxy formation
\citep{Nagetal04} are consistent with the stellar mass function around
$z=1$ and with the cosmic star formation history \citep[see
also][]{Fonetal04}, hence global quantities are well
reproduced. However, the models have difficulty producing the old
average ages of massive galaxies in the local universe, and
accordingly the number density of luminous red galaxies at
intermediate redshift as well as the high star formation rates in
massive objects at high redshifts \citep{Somerville04,Sometal04}.

At least part of the reason for this shortfall may be connected to the
way non-baryonic dark matter and baryonic matter are linked. In
current models, star formation is tightly linked to the assembly
history of dark matter halos, so that galaxies with longer assembly
times also form stars on longer timescales.  A possible solution to
reconcile extended assembly times with fast star formation histories
is a scenario, in which massive objects form their stars early and
assemble later \citep{BKT98,vDoketal99}. This picture gets some
theoretical support from the fact that the kinematics of massive
early-type galaxies require the origin of mergers between
bulge-dominated rather than disk galaxies \citep{NB03,KB03}. However,
in this case, it would be difficult for early-type galaxies to
establish the correlation between metallicity and mass reported
here. More fundamental modifications of the models are certainly
necessary.

\smallskip
\citet{Graetal01} have presented a promising---even though quite
heuristic---approach, in which star formation is enhanced in massive
systems because of feedback between star formation and AGN activity
(see also \citealt{CB03} and \citealt{KG04}). This idea is now put on
physical grounds in \citet{Graetal04}, where the evolution of gas in a
dark matter halo is described as a function of gravity, radiative
cooling, heating by feedback from supernovae, and---most
importantly---heating by feedback from the growing active nucleus. 

The assumption of a tight interplay between star formation and nuclear
activity gets indeed strong observational support from the fact that
besides the \aFe\ ratio, also black hole mass correlates with velocity
dispersion and galaxy mass \citep{Gebetal00,MF00}.  The resulting
'anti-hierarchical baryonic collapse' in the \citet{Graetal04} model
leads to higher formation redshifts and shorter formation timescales
of the stellar populations in massive halos. In this way both the
relationship between \aFe\ ratio and galaxy mass \citep{Rometal02},
and the presence of old galaxies at $z\sim 1$ (EROs) and of violently
star forming objects at $z\sim 3$ (SCUBA sources) are reproduced.

\citet{Graetal04} consider the evolution of massive halos in the
framework of semi-analytic models only above redshifts $z\ga 1$.
Hence, a fundamental problem that remains to be solved is the
accretion of cold gas onto the central galaxy of a dark matter halo at
lower redshift leading to late star formation and the galaxy's
rejuvenation.  

An important step to a solution of this latter problem might have been
recently made by \citet{Binney04}, who suggests a scenario in which
gas cooling is the key process. Unlike small systems with shallow
potential wells, massive galaxies cannot eject but retain their hot
gas, the cooling of which is inhibited by an episodically active
galactic nucleus. This 'atmosphere' of hot gas builds up a shield
around the galaxy preventing cool gas to fall in and to form
stars. The inclusion of this effect in semi-analytic models of
hierarchical galaxy formation would certainly be the right step toward
reconciling our current understanding of galaxy formation with the
observational constraints set by the stellar populations properties of
massive galaxies.

\section{Conclusions}
\label{sec:conclusions}
We study the stellar population properties of 124 early-type galaxies
\citep{G93,Beuetal02,Mehetal03} with velocity dispersions between 50
and 350$\;$km/s in both high and low density environments. From the
absorption line indices \Hb, \Mgb, and \Fe\ we derive ages,
metallicities \ZH, and \aFe\ element ratios using our abundance ratio
sensitive stellar population models (TMB).  The \aFe\ ratio is the
key to determine the formation timescales of the objects, as it
quantifies the relative importance of the chemical enrichment from
Type~II and the delayed Type~Ia supernovae. Together with average
ages, we use this information to set constraints on the epochs, when
early-type galaxies form the bulk of their stars. To get a handle on
the effect of correlated errors, we compare the observed sample with
mock galaxy samples with properties as close as possible to the
observational data in terms of sample size, velocity dispersion
distribution, and measurement errors.

For the bulk of the sample we find that all three stellar population
parameters age, total metallicity, and \aFe\ ratio correlate with
velocity dispersion, hence galaxy mass.  We show that these results
are robust against recent revisions of the local abundance pattern at
high metallicities.  To recover the observed scatter we need to assume
an intrinsic scatter of about 20 per cent in age, $0.08\;$dex in [\ZH]
and $0.05\;$dex in [\aFe].  Both zero-point and slope of the
\aFe-$\sigma$ relation are independent of the environmental density,
while the ages of objects in low density environments appear
systematically lower accompanied by slightly higher metallicities.

We identify three classes of 'outliers' from these relationships.  1)
The low-mass objects with $M_*\la 10^{10}\;\Msun$ ($\sigma<130\;$km/s)
in our sample show clear evidence for the presence of intermediate-age
populations. Their line indices are best reproduced with a 2-component
model in which a with a 1.5--2$\;$Gyr population with low \aFe\ ratios
is added to a base population defined by the above relationships. 2)
About 20 per cent of the intermediate-mass objects with
$10^{10}\la M_*/\Msun\la 10^{11}$ ($110\la\sigma/{\rm km/s}\la 230$)
must have either a young subpopulation or a blue horizontal
branch. Compared to the low-mass objects, the case for the
intermediate-age population is less convincing because of relatively
high \aFe\ ratios. 3) Most intriguing is a population of massive
($M_*\ga 10^{11}\;\Msun$ or $\sigma\ga 200\;$km/s) S0 galaxies, only
present in the high density sample, that require a major fraction of
their stars ($\sim 70$ per cent) to have formed relatively recently at
lookback times of about $2$--$3\;$Gyr, corresponding to $z\sim 0.2$.
Note that these objects, even though being interesting in themselves,
represent only a minor fraction of the total stellar mass budget in
early-type galaxies.

\medskip
The increase of total metallicity, indicating the completeness of
chemical processing, mainly reflects the inability of massive galaxies
with deeper potential wells to develop galactic winds. The increase of
age and \aFe\ ratio with $\sigma$ shows that more massive galaxies
form their stellar populations both earlier and faster. Our results
suggest that the well-known Mg-$\sigma$ relation is mainly driven by
metallicity ($\sim 60$ per cent contribution) with almost equal share
from \aFe\ ratio ($\sim 23$ per cent) and age ($\sim 17$ per cent).
The increase of \aFe\ with velocity dispersion is responsible for the
much weaker correlation of \Fe\ with galaxy mass.

With a simple chemical evolution model that takes the delayed
enrichment from Type~Ia supernovae into account, we translate the ages
and \aFe\ ratios of the above relationships into star formation
histories as a functions of galaxy's velocity dispersions and
environmental densities. We show that the higher ages and \aFe\ ratios
of more massive early-type galaxies imply faster formation timescales
at earlier epochs. The lower the mass of a galaxy is, the more it
suffers from episodes of late star formation extending to redshifts
below $z\sim 1$ for galaxies with $M_*\la 10^{10}\;\Msun$. Massive
early-type galaxies ($M_*\ga 10^{11}\;\Msun$) in high density
environments form their stellar populations between redshifts 2 and~5.
Galaxies in low density environments appear on average $1$--$2\;$Gyr
younger than their counterparts in high environmental densities, but
obey the same relationship between \aFe\ ratio and velocity
dispersion. This implies that they form on the same timescales, but
the epochs of their formation are delayed to lower redshifts between
$z\sim 1$ and $z\sim 2$.

\medskip 
The 'archaeology approach' pursued in this paper and observations of
the high-redshift universe are now converging to a consistent picture.
The results of this work imply that at least 50 per cent of the total
stellar mass density must have already been formed by $z\sim 1$, which
is in good agreement with observational estimates of the total stellar
mass density as a function of redshift. We conclude that significant
mass growth in the early-type galaxy population below $z\sim 1$ must
be restricted to less massive objects, and a significant increase of
the stellar mass density between redshifts 1 and 2 should be present
caused mainly by the field galaxy population.  As shown here, the
properties of local galaxies predict the presence of massive objects
with very high star formation rates of the order $1000\;$\Msun
yr$^{-1}$ around redshifts 2--5, and the existence of old evolved
galaxies at redshifts between 1 and 2, both now confirmed
observationally.

Current models of hierarchical galaxy formation are consistent with
global properties like the stellar mass function, but remain
challenged by the ages of local massive galaxies , the number
densities of luminous red objects at intermediate redshift, and the
high star formation rates in massive objects at high redshifts. The
problem seems tightly linked to the (uncertain) treatment of baryonic
matter in dark halos, and likely solutions lie in the inclusion of AGN
feedback and refinement of gas heating/cooling processes.

\acknowledgments We thank the referee, Robert Proctor, for the careful
reading of the manuscript and for his very constructive report.



\appendix
\section{Data sample}
In Table~\ref{tab:sample} we list list velocity dispersions, line
indices \Hb, \Mgb, and \Fe, galaxy types, environmental densities,
literature sources, and classifications of Fig.~\ref{fig:mixgrid} of
the sample investigated. Table~\ref{tab:results} gives the stellar
population parameters age, metallicity, and \aFe\ ratio derived with
the TMB models. The 1-$\sigma$ errors are estimated through Monte
Carlo simulations.  We perturb the line indices assuming a Gaussian
probability distribution defined by the 1-$\sigma$ measurement errors
quoted above.  The resulting 1-$\sigma$ errors correspond to the
widths of the distributions in the stellar population parameters from
100 Monte Carlo realization per object.

\begin{deluxetable}{lrccccccccccc}
\tablecaption{Sample parameters and line indices\label{tab:sample}}
\tablewidth{0pt}
\tablehead{\colhead{Name} & \colhead{Sigma} & \colhead{Error} & \colhead{H$\beta$} & \colhead{Error} & \colhead{Mg$\, b$} & \colhead{Error} & \colhead{$\langle {\rm Fe}\rangle$} & \colhead{Error} & \colhead{Type} & \colhead{Environment} & \colhead{Source} & \colhead{Class}}
\startdata
%
     NGC 0221   &     72.1  &    2.6 &  2.31 &  0.05 &  2.96 &  0.03 &  2.75 &  0.03    &      E &      Low density  &  Gonzalez & triangles\\
     NGC 0224   &    156.1  &    3.7 &  1.67 &  0.07 &  4.85 &  0.05 &  3.10 &  0.04    &     S0 &      Low density  &  Gonzalez &   circles\\
     NGC 0315   &    321.0  &    3.9 &  1.74 &  0.06 &  4.84 &  0.05 &  2.88 &  0.05    &      E &      Low density  &  Gonzalez &   circles\\
     NGC 0507   &    262.2  &    6.4 &  1.73 &  0.09 &  4.52 &  0.11 &  2.78 &  0.10    &      E &      Low density  &  Gonzalez &   circles\\
     NGC 0547   &    235.6  &    3.6 &  1.58 &  0.07 &  5.02 &  0.05 &  2.82 &  0.05    &      E &      Low density  &  Gonzalez &   circles\\
     NGC 0584   &    193.2  &    2.8 &  2.08 &  0.05 &  4.33 &  0.04 &  2.90 &  0.03    &     S0 &      Low density  &  Gonzalez &   squares\\
     NGC 0636   &    160.3  &    2.9 &  1.89 &  0.04 &  4.20 &  0.04 &  3.03 &  0.04    &      E &      Low density  &  Gonzalez &   circles\\
     NGC 0720   &    238.6  &    5.2 &  1.77 &  0.12 &  5.17 &  0.11 &  2.87 &  0.09    &      E &      Low density  &  Gonzalez &   circles\\
     NGC 0821   &    188.7  &    2.9 &  1.66 &  0.04 &  4.53 &  0.04 &  2.95 &  0.04    &      E &      Low density  &  Gonzalez &   circles\\
     NGC 1453   &    286.5  &    3.5 &  1.60 &  0.06 &  4.95 &  0.05 &  2.98 &  0.05    &      E &      Low density  &  Gonzalez &   circles\\
     NGC 1600   &    314.8  &    4.3 &  1.55 &  0.07 &  5.13 &  0.06 &  3.06 &  0.06    &      E &      Low density  &  Gonzalez &   circles\\
     NGC 1700   &    227.3  &    3.0 &  2.11 &  0.05 &  4.15 &  0.04 &  3.00 &  0.04    &      E &      Low density  &  Gonzalez &   squares\\
     NGC 2300   &    251.8  &    3.3 &  1.68 &  0.06 &  4.98 &  0.05 &  2.97 &  0.05    &      E &      Low density  &  Gonzalez &   circles\\
     NGC 2778   &    154.4  &    3.2 &  1.77 &  0.08 &  4.70 &  0.06 &  2.85 &  0.05    &      E &      Low density  &  Gonzalez &   circles\\
     NGC 3377   &    107.6  &    2.6 &  2.09 &  0.05 &  3.99 &  0.03 &  2.61 &  0.03    &      E &      Low density  &  Gonzalez &   squares\\
     NGC 3379   &    203.2  &    2.7 &  1.62 &  0.05 &  4.78 &  0.03 &  2.86 &  0.03    &      E &      Low density  &  Gonzalez &   circles\\
     NGC 3608   &    177.7  &    3.0 &  1.69 &  0.06 &  4.61 &  0.04 &  2.94 &  0.04    &      E &      Low density  &  Gonzalez &   circles\\
     NGC 3818   &    173.2  &    4.2 &  1.71 &  0.08 &  4.88 &  0.07 &  2.97 &  0.06    &      E &      Low density  &  Gonzalez &   circles\\
     NGC 4261   &    288.3  &    3.3 &  1.34 &  0.06 &  5.11 &  0.04 &  3.01 &  0.04    &      E &    High density  &  Gonzalez &   circles\\
     NGC 4278   &    232.5  &    2.8 &  1.56 &  0.05 &  4.92 &  0.04 &  2.68 &  0.04    &      E &      Low density  &  Gonzalez &   circles\\
     NGC 4374   &    282.1  &    2.8 &  1.51 &  0.04 &  4.78 &  0.03 &  2.82 &  0.03    &      E &    High density  &  Gonzalez &   circles\\
     NGC 4472   &    279.2  &    4.4 &  1.62 &  0.06 &  4.85 &  0.06 &  2.91 &  0.05    &      E &    High density  &  Gonzalez &   circles\\
     NGC 4478   &    127.7  &    4.4 &  1.84 &  0.06 &  4.33 &  0.06 &  2.94 &  0.05    &      E &    High density  &  Gonzalez &   circles\\
     NGC 4489   &     47.2  &    3.8 &  2.39 &  0.07 &  3.21 &  0.06 &  2.66 &  0.05    &      E &    High density  &  Gonzalez & triangles\\
     NGC 4552   &    251.8  &    2.8 &  1.47 &  0.05 &  5.15 &  0.03 &  2.99 &  0.03    &      E &    High density  &  Gonzalez &   circles\\
     NGC 4649   &    309.8  &    3.0 &  1.40 &  0.05 &  5.33 &  0.04 &  3.01 &  0.03    &     S0 &    High density  &  Gonzalez &   circles\\
     NGC 4697   &    162.4  &    3.9 &  1.75 &  0.07 &  4.08 &  0.05 &  2.77 &  0.04    &      E &    High density  &  Gonzalez &   circles\\
     NGC 5638   &    154.2  &    2.8 &  1.65 &  0.04 &  4.64 &  0.04 &  2.84 &  0.04    &      E &      Low density  &  Gonzalez &   circles\\
     NGC 5812   &    200.3  &    3.0 &  1.70 &  0.04 &  4.81 &  0.04 &  3.06 &  0.04    &      E &      Low density  &  Gonzalez &   circles\\
     NGC 5813   &    204.8  &    3.2 &  1.42 &  0.07 &  4.65 &  0.05 &  2.67 &  0.05    &      E &      Low density  &  Gonzalez &   circles\\
     NGC 5831   &    160.5  &    2.7 &  2.00 &  0.05 &  4.38 &  0.04 &  3.05 &  0.03    &      E &      Low density  &  Gonzalez &  diamonds\\
     NGC 5846   &    223.5  &    4.0 &  1.45 &  0.07 &  4.93 &  0.05 &  2.86 &  0.04    &     S0 &      Low density  &  Gonzalez &   circles\\
     NGC 6127   &    238.9  &    4.4 &  1.50 &  0.05 &  4.96 &  0.06 &  2.85 &  0.05    &      E &      Low density  &  Gonzalez &   circles\\
     NGC 6702   &    173.8  &    3.0 &  2.46 &  0.06 &  3.80 &  0.04 &  3.00 &  0.04    &      E &      Low density  &  Gonzalez &   squares\\
     NGC 6703   &    182.8  &    2.9 &  1.88 &  0.06 &  4.30 &  0.04 &  2.93 &  0.04    &     S0 &      Low density  &  Gonzalez &   circles\\
     NGC 7052   &    273.8  &    3.6 &  1.48 &  0.07 &  5.02 &  0.06 &  2.84 &  0.05    &      E &      Low density  &  Gonzalez &   circles\\
     NGC 7454   &    106.5  &    2.9 &  2.15 &  0.06 &  3.27 &  0.05 &  2.48 &  0.04    &      E &      Low density  &  Gonzalez & triangles\\
     NGC 7562   &    248.0  &    2.9 &  1.69 &  0.05 &  4.54 &  0.04 &  2.87 &  0.04    &      E &    High density  &  Gonzalez &   circles\\
     NGC 7619   &    300.3  &    3.1 &  1.36 &  0.04 &  5.06 &  0.04 &  3.06 &  0.04    &      E &    High density  &  Gonzalez &   circles\\
     NGC 7626   &    253.1  &    3.0 &  1.46 &  0.05 &  5.05 &  0.04 &  2.83 &  0.04    &      E &    High density  &  Gonzalez &   circles\\
     NGC 7785   &    239.6  &    3.1 &  1.63 &  0.06 &  4.60 &  0.04 &  2.91 &  0.04    &      E &      Low density  &  Gonzalez &   circles\\
       NGC 4874 &    259.6  &    4.3 &  2.13 &  0.02 &  4.93 &  0.02 &  2.91 &  0.02    &     cD &    High density  &   Mehlert &  diamonds\\
       NGC 4889 &    358.7  &    4.2 &  1.93 &  0.03 &  5.45 &  0.04 &  3.04 &  0.03    &     cD &    High density  &   Mehlert &  diamonds\\
       NGC 4839 &    275.5  &    5.8 &  1.42 &  0.04 &  4.92 &  0.04 &  2.75 &  0.04    &      E &    High density  &   Mehlert &   circles\\
      NGC 4841A &    263.9  &    7.5 &  1.53 &  0.05 &  4.51 &  0.05 &  2.89 &  0.04    &      E &    High density  &   Mehlert &   circles\\
       NGC 4926 &    273.3  &    4.6 &  1.50 &  0.06 &  5.17 &  0.06 &  2.50 &  0.05    &      E &    High density  &   Mehlert &   circles\\
       IC 4051  &    258.7  &    2.1 &  1.42 &  0.06 &  5.34 &  0.07 &  2.75 &  0.05    &      E &    High density  &   Mehlert &   circles\\
       NGC 4895 &    207.2  &    1.2 &  1.68 &  0.05 &  4.39 &  0.06 &  2.60 &  0.05    &     S0 &    High density  &   Mehlert &   circles\\
       NGC 4860 &    280.5  &    4.1 &  1.39 &  0.06 &  5.39 &  0.07 &  2.85 &  0.05    &      E &    High density  &   Mehlert &   circles\\
       NGC 4896 &    209.1  &    4.8 &  1.72 &  0.06 &  4.04 &  0.06 &  2.67 &  0.05    &     S0 &    High density  &   Mehlert &   circles\\
       NGC 4865 &    249.7  &    4.1 &  2.26 &  0.08 &  4.51 &  0.09 &  3.01 &  0.07    &     S0 &    High density  &   Mehlert &  diamonds\\
       NGC 4923 &    186.0  &    1.9 &  1.70 &  0.05 &  4.43 &  0.05 &  2.69 &  0.04    &      E &    High density  &   Mehlert &   circles\\
       GMP 2413 &    180.0  &    3.0 &  1.80 &  0.06 &  4.08 &  0.06 &  2.60 &  0.05    &     S0 &    High density  &   Mehlert &   circles\\
       NGC 4840 &    216.6  &    2.7 &  1.63 &  0.07 &  4.94 &  0.07 &  2.91 &  0.06    &      E &    High density  &   Mehlert &   circles\\
       NGC 4869 &    188.1  &    2.1 &  1.40 &  0.05 &  4.83 &  0.05 &  2.90 &  0.04    &      E &    High density  &   Mehlert &   circles\\
       NGC 4908 &    192.4  &    2.7 &  1.58 &  0.09 &  4.58 &  0.09 &  2.65 &  0.07    &      E &    High density  &   Mehlert &   circles\\
       IC 4045  &    167.3  &    2.1 &  1.46 &  0.06 &  4.70 &  0.07 &  2.77 &  0.05    &      E &    High density  &   Mehlert &   circles\\
       NGC 4871 &    154.3  &    3.0 &  1.61 &  0.10 &  4.36 &  0.11 &  2.61 &  0.09    &     S0 &    High density  &   Mehlert &   circles\\
       NGC 4850 &    155.8  &    2.0 &  1.57 &  0.06 &  4.39 &  0.06 &  2.58 &  0.05    &      E &    High density  &   Mehlert &   circles\\
       NGC 4883 &    157.9  &    2.0 &  1.58 &  0.06 &  4.44 &  0.06 &  2.86 &  0.05    &     S0 &    High density  &   Mehlert &   circles\\
       GMP 1853  &    160.4  &    5.0 &  1.63 &  0.11 &  4.26 &  0.12 &  2.92 &  0.09    &     S0 &    High density  &   Mehlert &   circles\\
       GMP 3661  &    154.2  &    4.1 &  2.25 &  0.08 &  4.20 &  0.08 &  2.83 &  0.07    &     S0 &    High density  &   Mehlert &   squares\\
       GMP 4679  &     82.2  &    2.5 &  2.39 &  0.08 &  3.31 &  0.09 &  2.50 &  0.07    &     S0 &    High density  &   Mehlert & triangles\\
       NGC 4872 &    171.7  &    3.1 &  2.05 &  0.05 &  4.05 &  0.06 &  2.82 &  0.04    &      E &    High density  &   Mehlert &   squares\\
       IC 4041  &    114.2  &    3.2 &  2.19 &  0.08 &  3.82 &  0.09 &  2.82 &  0.07    &     S0 &    High density  &   Mehlert &   squares\\
       NGC 4957 &    208.4  &    2.0 &  1.76 &  0.03 &  4.53 &  0.03 &  2.93 &  0.02    &      E &    High density  &   Mehlert &   circles\\
       NGC 4952 &    252.6  &    1.7 &  1.71 &  0.03 &  4.76 &  0.03 &  2.69 &  0.02    &      E &    High density  &   Mehlert &   circles\\
       NGC 4944 &    144.8  &    1.1 &  2.13 &  0.03 &  3.83 &  0.03 &  2.64 &  0.02    &     S0 &    High density  &   Mehlert &   squares\\
       NGC 4931 &    150.8  &    1.2 &  2.10 &  0.04 &  3.88 &  0.05 &  2.78 &  0.04    &     S0 &    High density  &   Mehlert &   squares\\
       GMP 1990   &    208.9  &    2.4 &  1.40 &  0.04 &  4.78 &  0.04 &  2.50 &  0.03    &      E &    High density  &   Mehlert &   circles\\
       NGC 4827 &    243.7  &    2.1 &  1.53 &  0.03 &  4.89 &  0.03 &  2.80 &  0.02    &      E &    High density  &   Mehlert &   circles\\
       NGC 4816 &    231.1  &    2.4 &  1.61 &  0.02 &  4.87 &  0.02 &  2.74 &  0.02    &     S0 &    High density  &   Mehlert &   circles\\
       NGC 4807 &    178.5  &    2.0 &  1.81 &  0.06 &  4.39 &  0.06 &  2.78 &  0.05    &      E &    High density  &   Mehlert &   circles\\
\enddata 
\end{deluxetable}

\setcounter{table}{1}
\begin{deluxetable}{lrccccccccccc}
\tablecaption{-- continued.}
\tablewidth{0pt}
\tablehead{\colhead{Name} & \colhead{Sigma} & \colhead{Error} & \colhead{H$\beta$} & \colhead{Error} & \colhead{Mg$\, b$} & \colhead{Error} & \colhead{$\langle {\rm Fe}\rangle$} & \colhead{Error} & \colhead{Type} & \colhead{Environment} & \colhead{Source} & \colhead{Class}}
\startdata
%
 ESO 244G045   &    185.1  &    3.7 &  1.66 &  0.09 &  4.12 &  0.08 &  2.86 &  0.05    &     S0 &      Low density  &    Beuing &   circles\\
 ESO 1070040   &    147.0  &    5.8 &  2.24 &  0.25 &  3.63 &  0.16 &  2.97 &  0.09    &      E &      Low density  &    Beuing &   squares\\
 ESO 1370080   &    297.3  &    7.1 &  1.62 &  0.18 &  5.16 &  0.15 &  3.18 &  0.09    &     S0 &    High density  &    Beuing &  diamonds\\
 ESO 1480170   &    134.5  &    7.5 &  2.26 &  0.52 &  3.49 &  0.32 &  2.58 &  0.20    &      E &      Low density  &    Beuing & triangles\\
 ESO 1850540   &    277.2  &    4.3 &  1.57 &  0.06 &  5.11 &  0.05 &  3.07 &  0.03    &      E &    High density  &    Beuing &   circles\\
 ESO 2210200   &    153.0  &    5.0 &  1.90 &  0.17 &  4.01 &  0.13 &  2.83 &  0.08    &     S0 &      Low density  &    Beuing &   circles\\
 ESO 3220600   &    193.2  &    3.8 &  1.99 &  0.17 &  4.45 &  0.14 &  3.08 &  0.08    &     S0 &    High density  &    Beuing &  diamonds\\
 ESO 4230240   &    182.5  &    5.5 &  2.06 &  0.37 &  3.84 &  0.24 &  2.97 &  0.14    &     S0 &      Low density  &  Oliveira &   squares\\
 ESO 4430240   &    287.0  &    7.8 &  1.75 &  0.15 &  5.15 &  0.13 &  3.13 &  0.08    &     S0 &    High density  &    Beuing &  diamonds\\
 ESO 4450020   &    255.5  &    6.4 &  1.69 &  0.09 &  4.76 &  0.08 &  3.02 &  0.05    &     S0 &      Low density  &    Beuing &   circles\\
 ESO 4940350   &    132.1  &    2.4 &  2.20 &  0.16 &  3.53 &  0.12 &  2.74 &  0.08    &     S0 &      Low density  &  Oliveira & triangles\\
      IC 1633   &    347.7  &    7.9 &  1.73 &  0.09 &  5.42 &  0.08 &  2.98 &  0.05    &     S0 &    High density  &    Beuing &  diamonds\\
      IC 3152   &    167.8  &    3.7 &  1.82 &  0.19 &  3.66 &  0.16 &  2.89 &  0.10    &     S0 &      Low density  &    Beuing &   circles\\
      IC 3986   &    311.5  &    7.0 &  1.87 &  0.15 &  5.03 &  0.12 &  3.12 &  0.08    &     S0 &    High density  &    Beuing &  diamonds\\
      IC 4797   &    220.6  &    6.4 &  1.92 &  0.26 &  4.52 &  0.18 &  2.75 &  0.10    &      E &      Low density  &    Beuing &   circles\\
      IC 4931   &    289.0  &    6.0 &  1.80 &  0.08 &  4.76 &  0.07 &  3.06 &  0.04    &     S0 &    High density  &    Beuing &  diamonds\\
      IC 5105   &    308.1  &    5.4 &  1.57 &  0.06 &  5.26 &  0.05 &  2.37 &  0.03    &     S0 &      Low density  &  Oliveira &   circles\\
      NGC 0312  &    254.8  &    4.4 &  1.83 &  0.09 &  4.56 &  0.08 &  2.48 &  0.05    &      E &      Low density  &    Beuing &   circles\\
      NGC 0484  &    244.5  &    4.5 &  1.60 &  0.08 &  4.41 &  0.07 &  2.95 &  0.04    &     S0 &      Low density  &    Beuing &   circles\\
      NGC 0596  &    161.8  &    1.6 &  2.12 &  0.05 &  3.95 &  0.04 &  2.81 &  0.03    &      E &      Low density  &  Oliveira &   squares\\
      NGC 0636  &    178.5  &    5.0 &  1.86 &  0.26 &  4.38 &  0.17 &  2.83 &  0.09    &      E &      Low density  &  Oliveira &   circles\\
      NGC 0731  &    162.1  &    2.6 &  2.10 &  0.08 &  3.92 &  0.07 &  2.86 &  0.04    &     S0 &      Low density  &    Beuing &   squares\\
      NGC 1052  &    202.6  &    1.7 &  1.22 &  0.04 &  5.53 &  0.03 &  2.77 &  0.02    &      E &      Low density  &    Beuing &   circles\\
      NGC 1172  &    119.3  &    1.5 &  1.82 &  0.11 &  3.70 &  0.08 &  2.52 &  0.06    &     S0 &      Low density  &  Oliveira &   circles\\
      NGC 1316  &    223.1  &    2.1 &  2.07 &  0.03 &  3.90 &  0.03 &  2.84 &  0.02    &     S0 &    High density  &  Oliveira &   squares\\
      NGC 1340  &    175.4  &    1.5 &  2.23 &  0.05 &  3.99 &  0.04 &  3.02 &  0.03    &     S0 &    High density  &  Oliveira &   squares\\
      NGC 1375  &     75.1  &    1.5 &  2.35 &  0.53 &  2.68 &  0.33 &  2.56 &  0.20    &     S0 &    High density  &    Beuing & triangles\\
      NGC 1395  &    250.0  &    3.0 &  1.62 &  0.05 &  5.21 &  0.04 &  2.93 &  0.03    &      E &      Low density  &  Oliveira &   circles\\
      NGC 1400  &    256.2  &    3.0 &  1.38 &  0.05 &  5.16 &  0.04 &  2.93 &  0.03    &     S0 &      Low density  &  Oliveira &   circles\\
      NGC 1407  &    259.7  &    3.7 &  1.67 &  0.07 &  4.88 &  0.06 &  2.85 &  0.03    &      E &      Low density  &  Oliveira &   circles\\
      NGC 1411  &    144.0  &    0.9 &  1.74 &  0.04 &  3.90 &  0.04 &  2.57 &  0.03    &     S0 &      Low density  &  Oliveira &   circles\\
      NGC 1549  &    203.3  &    1.4 &  1.79 &  0.03 &  4.39 &  0.03 &  2.88 &  0.02    &      E &      Low density  &  Oliveira &   circles\\
      NGC 1553  &    207.2  &    1.1 &  1.88 &  0.03 &  4.44 &  0.03 &  3.09 &  0.02    &     S0 &      Low density  &  Oliveira &   circles\\
      NGC 2434  &    180.4  &    3.9 &  1.87 &  0.13 &  3.72 &  0.10 &  2.87 &  0.07    &      E &      Low density  &    Beuing &   circles\\
      NGC 2717  &    210.0  &    6.5 &  1.91 &  0.18 &  4.69 &  0.14 &  2.86 &  0.09    &     S0 &      Low density  &  Oliveira &   circles\\
      NGC 2891  &    100.8  &    0.6 &  2.54 &  0.20 &  3.07 &  0.16 &  2.33 &  0.11    &     S0 &      Low density  &    Beuing & triangles\\
      NGC 2986  &    282.2  &    4.2 &  1.48 &  0.06 &  4.97 &  0.05 &  2.92 &  0.03    &      E &      Low density  &  Oliveira &   circles\\
      NGC 3078  &    268.1  &    4.2 &  1.12 &  0.09 &  5.20 &  0.07 &  3.16 &  0.04    &      E &      Low density  &  Oliveira &   circles\\
      NGC 3224  &    155.8  &    2.3 &  2.31 &  0.14 &  3.91 &  0.12 &  2.92 &  0.08    &      E &    High density  &    Beuing &   squares\\
      NGC 3923  &    267.9  &    3.9 &  1.87 &  0.08 &  5.12 &  0.07 &  3.07 &  0.04    &      E &      Low density  &    Beuing &  diamonds\\
      NGC 4786  &    291.6  &    5.8 &  1.61 &  0.09 &  4.57 &  0.08 &  2.87 &  0.05    &     S0 &      Low density  &    Beuing &   circles\\
      NGC 5087  &    303.3  &    5.1 &  1.62 &  0.11 &  5.39 &  0.09 &  3.06 &  0.05    &     S0 &      Low density  &    Beuing &   circles\\
      NGC 5516  &    313.0  &   10.1 &  1.46 &  0.08 &  4.81 &  0.06 &  3.29 &  0.04    &     S0 &      Low density  &    Beuing &   circles\\
      NGC 5791  &    271.8  &    9.8 &  1.60 &  0.19 &  5.06 &  0.15 &  3.30 &  0.10    &      E &      Low density  &    Beuing &   circles\\
      NGC 5903  &    209.2  &    3.7 &  1.68 &  0.10 &  4.44 &  0.08 &  2.90 &  0.05    &      E &      Low density  &    Beuing &   circles\\
      NGC 6305  &    158.9  &    4.3 &  2.03 &  0.13 &  4.07 &  0.10 &  2.80 &  0.07    &     S0 &      Low density  &    Beuing &   squares\\
      NGC 6958  &    193.9  &    2.1 &  1.63 &  0.05 &  3.89 &  0.04 &  2.75 &  0.03    &     S0 &      Low density  &    Beuing &   circles\\
      NGC 7041  &    238.7  &    1.0 &  1.88 &  0.03 &  4.70 &  0.03 &  2.52 &  0.02    &     S0 &      Low density  &  Oliveira &   circles\\
      NGC 7185  &    101.2  &    1.7 &  2.35 &  0.16 &  3.05 &  0.13 &  2.63 &  0.09    &     S0 &      Low density  &    Beuing & triangles\\
      NGC 7365  &    115.4  &    5.5 &  2.20 &  0.47 &  3.25 &  0.30 &  2.56 &  0.18    &     S0 &      Low density  &    Beuing & triangles\\
      NGC 7796  &    264.1  &    5.5 &  1.51 &  0.08 &  5.02 &  0.07 &  2.78 &  0.04    &     S0 &      Low density  &  Oliveira &   circles
\enddata 
\tablecomments{Cols.~3, 5, 7, 9 give 1-$\sigma$ errors. Sources for
the data (col.~12): \citet{G93}, \citet{Beuetal02}, \citet{Mehetal03},
C.~Mendes de Oliveira et al.\ (in preparation). The last column
indicates the classification shown by symbol type in
Fig.~\ref{fig:mixgrid}.}
\end{deluxetable}

\begin{deluxetable}{lrccccc}
\tablecaption{Results\label{tab:results}}
\tablewidth{0pt}
\tablehead{\colhead{Name} & \colhead{Age} & \colhead{Error} & \colhead{[\ZH]} & \colhead{Error} & \colhead{[\aFe]} & \colhead{Error}}
\startdata
%
     NGC 0221   &     2.4  &    0.2   & 0.152 &  0.030  &-0.025  & 0.013 \\
     NGC 0224   &     7.0  &    1.7   & 0.441 &  0.048  & 0.219  & 0.017 \\
     NGC 0315   &     6.6  &    1.4   & 0.397 &  0.038  & 0.282  & 0.019 \\
     NGC 0507   &     8.1  &    1.9   & 0.261 &  0.071  & 0.246  & 0.040 \\
     NGC 0547   &    10.7  &    1.7   & 0.321 &  0.049  & 0.311  & 0.020 \\
     NGC 0584   &     2.8  &    0.3   & 0.478 &  0.046  & 0.223  & 0.014 \\
     NGC 0636   &     4.4  &    0.6   & 0.376 &  0.028  & 0.133  & 0.013 \\
     NGC 0720   &     5.4  &    2.4   & 0.485 &  0.083  & 0.353  & 0.039 \\
     NGC 0821   &     8.9  &    1.2   & 0.297 &  0.036  & 0.189  & 0.015 \\
     NGC 1453   &     9.4  &    1.6   & 0.380 &  0.044  & 0.256  & 0.018 \\
     NGC 1600   &     9.7  &    1.9   & 0.420 &  0.046  & 0.263  & 0.021 \\
     NGC 1700   &     2.6  &    0.3   & 0.500 &  0.049  & 0.167  & 0.014 \\
     NGC 2300   &     7.3  &    1.5   & 0.427 &  0.040  & 0.279  & 0.017 \\
     NGC 2778   &     6.4  &    1.9   & 0.366 &  0.049  & 0.268  & 0.023 \\
     NGC 3377   &     3.6  &    0.5   & 0.253 &  0.035  & 0.234  & 0.014 \\
     NGC 3379   &    10.0  &    1.1   & 0.299 &  0.036  & 0.259  & 0.012 \\
     NGC 3608   &     8.0  &    1.5   & 0.334 &  0.042  & 0.213  & 0.016 \\
     NGC 3818   &     6.7  &    1.9   & 0.422 &  0.053  & 0.263  & 0.025 \\
     NGC 4261   &    16.3  &    1.8   & 0.275 &  0.035  & 0.255  & 0.015 \\
     NGC 4278   &    12.0  &    1.3   & 0.225 &  0.038  & 0.331  & 0.015 \\
     NGC 4374   &    12.8  &    1.1   & 0.207 &  0.032  & 0.259  & 0.011 \\
     NGC 4472   &     9.6  &    1.4   & 0.342 &  0.046  & 0.258  & 0.021 \\
     NGC 4478   &     5.3  &    1.2   & 0.344 &  0.049  & 0.180  & 0.021 \\
     NGC 4489   &     2.0  &    0.3   & 0.225 &  0.043  & 0.078  & 0.021 \\
     NGC 4552   &    12.4  &    1.5   & 0.356 &  0.034  & 0.277  & 0.011 \\
     NGC 4649   &    14.1  &    1.5   & 0.362 &  0.029  & 0.296  & 0.012 \\
     NGC 4697   &     8.3  &    1.4   & 0.148 &  0.043  & 0.155  & 0.018 \\
     NGC 5638   &     9.5  &    0.9   & 0.274 &  0.033  & 0.239  & 0.015 \\
     NGC 5812   &     6.5  &    1.0   & 0.432 &  0.029  & 0.226  & 0.014 \\
     NGC 5813   &    16.6  &    2.2   & 0.059 &  0.042  & 0.270  & 0.018 \\
     NGC 5831   &     3.0  &    0.4   & 0.504 &  0.043  & 0.186  & 0.017 \\
     NGC 5846   &    14.2  &    2.2   & 0.226 &  0.051  & 0.270  & 0.017 \\
     NGC 6127   &    12.6  &    1.5   & 0.264 &  0.044  & 0.284  & 0.017 \\
     NGC 6702   &     1.7  &    0.1   & 0.686 &  0.070  & 0.153  & 0.017 \\
     NGC 6703   &     4.8  &    0.8   & 0.354 &  0.034  & 0.180  & 0.016 \\
     NGC 7052   &    13.2  &    2.1   & 0.263 &  0.050  & 0.296  & 0.021 \\
     NGC 7454   &     3.7  &    0.6   & 0.032 &  0.044  & 0.117  & 0.021 \\
     NGC 7562   &     8.6  &    1.3   & 0.282 &  0.039  & 0.219  & 0.014 \\
     NGC 7619   &    15.4  &    1.4   & 0.291 &  0.031  & 0.233  & 0.013 \\
     NGC 7626   &    13.9  &    1.5   & 0.251 &  0.038  & 0.302  & 0.013 \\
     NGC 7785   &     9.7  &    1.6   & 0.280 &  0.049  & 0.209  & 0.016 \\
       NGC 4874 &     2.4  &    0.1   & 0.719 &  0.026  & 0.353  & 0.008 \\
       NGC 4889 &     2.8  &    0.1   & 0.769 &  0.043  & 0.388  & 0.020 \\
       NGC 4839 &    15.8  &    1.4   & 0.159 &  0.044  & 0.300  & 0.027 \\
      NGC 4841A &    12.2  &    1.3   & 0.178 &  0.049  & 0.186  & 0.024 \\
       NGC 4926 &    14.4  &    1.7   & 0.177 &  0.059  & 0.428  & 0.037 \\
       IC 4051  &    15.0  &    1.9   & 0.276 &  0.057  & 0.374  & 0.032 \\
       NGC 4895 &    10.4  &    1.3   & 0.115 &  0.054  & 0.266  & 0.032 \\
       NGC 4860 &    15.2  &    2.4   & 0.315 &  0.057  & 0.350  & 0.038 \\
       NGC 4896 &     9.7  &    1.4   & 0.073 &  0.056  & 0.170  & 0.032 \\
       NGC 4865 &     2.0  &    0.2   & 0.775 &  0.114  & 0.269  & 0.035 \\
       NGC 4923 &     9.4  &    1.1   & 0.180 &  0.055  & 0.249  & 0.037 \\
       GMP 2413 &     8.0  &    1.5   & 0.105 &  0.061  & 0.214  & 0.039 \\
       NGC 4840 &     9.2  &    1.8   & 0.371 &  0.058  & 0.273  & 0.030 \\
       NGC 4869 &    15.4  &    1.7   & 0.190 &  0.055  & 0.236  & 0.034 \\
       NGC 4908 &    12.2  &    2.5   & 0.127 &  0.084  & 0.278  & 0.052 \\
       IC 4045  &    14.6  &    2.1   & 0.134 &  0.075  & 0.255  & 0.044 \\
       NGC 4871 &    11.9  &    2.7   & 0.069 &  0.103  & 0.248  & 0.070 \\
       NGC 4850 &    13.1  &    1.6   & 0.037 &  0.068  & 0.257  & 0.055 \\
       NGC 4883 &    11.2  &    1.9   & 0.182 &  0.078  & 0.184  & 0.037 \\
       GMP 1853 &    10.2  &    3.0   & 0.185 &  0.112  & 0.130  & 0.065 \\
       GMP 3661 &     2.1  &    0.3   & 0.543 &  0.107  & 0.244  & 0.040 \\
       GMP 4679 &     2.2  &    0.3   & 0.183 &  0.058  & 0.150  & 0.041 \\
       NGC 4872 &     3.5  &    0.5   & 0.341 &  0.064  & 0.178  & 0.039 \\
       IC 4041  &     2.5  &    0.4   & 0.358 &  0.070  & 0.147  & 0.031 \\
       NGC 4957 &     6.5  &    0.7   & 0.351 &  0.022  & 0.210  & 0.011 \\
       NGC 4952 &     8.8  &    0.8   & 0.279 &  0.036  & 0.316  & 0.021 \\
       NGC 4944 &     3.3  &    0.3   & 0.250 &  0.033  & 0.195  & 0.020 \\
       NGC 4931 &     3.3  &    0.4   & 0.306 &  0.042  & 0.159  & 0.023 \\
       GMP 1990 &    18.2  &    1.6   & 0.011 &  0.042  & 0.349  & 0.038 \\
       NGC 4827 &    12.0  &    0.9   & 0.250 &  0.041  & 0.287  & 0.025 \\
       NGC 4816 &    10.6  &    0.5   & 0.267 &  0.030  & 0.311  & 0.020 \\
       NGC 4807 &     6.4  &    1.5   & 0.274 &  0.081  & 0.232  & 0.048 \\
\enddata 
\end{deluxetable}

\setcounter{table}{2}
\begin{deluxetable}{lrccccc}
\tablecaption{-- continued.}
\tablewidth{0pt}
\tablehead{\colhead{Name} & \colhead{Age} & \colhead{Error} & \colhead{[\ZH]} & \colhead{Error} & \colhead{[\aFe]} & \colhead{Error}}
\startdata
%
  ESO 244G045   &     9.9  &    2.1   & 0.141 &  0.064  & 0.121  & 0.025 \\
  ESO 1070040   &     2.1  &    2.3   & 0.414 &  0.190  & 0.081  & 0.059 \\
  ESO 1370080   &     7.0  &    4.0   & 0.517 &  0.130  & 0.251  & 0.042 \\
  ESO 1480170   &     2.7  &    4.0   & 0.196 &  0.189  & 0.153  & 0.088 \\
  ESO 1850540   &     9.2  &    1.5   & 0.434 &  0.035  & 0.259  & 0.014 \\
  ESO 2210200   &     5.1  &    2.5   & 0.246 &  0.112  & 0.148  & 0.045 \\
  ESO 3220600   &     3.0  &    1.7   & 0.533 &  0.154  & 0.192  & 0.044 \\
  ESO 4230240   &     3.3  &    4.7   & 0.324 &  0.220  & 0.125  & 0.083 \\
  ESO 4430240   &     4.4  &    2.4   & 0.575 &  0.107  & 0.277  & 0.040 \\
  ESO 4450020   &     7.2  &    2.0   & 0.402 &  0.060  & 0.223  & 0.026 \\
  ESO 4940350   &     2.8  &    1.2   & 0.226 &  0.097  & 0.150  & 0.043 \\
      IC 1633   &     5.2  &    1.6   & 0.563 &  0.057  & 0.366  & 0.024 \\
      IC 3152   &     6.4  &    3.6   & 0.132 &  0.131  & 0.034  & 0.051 \\
      IC 3986   &     3.3  &    2.0   & 0.623 &  0.131  & 0.280  & 0.039 \\
      IC 4797   &     4.7  &    4.2   & 0.362 &  0.181  & 0.279  & 0.067 \\
      IC 4931   &     4.7  &    1.5   & 0.475 &  0.057  & 0.230  & 0.023 \\
      IC 5105   &    11.4  &    1.7   & 0.210 &  0.041  & 0.481  & 0.013 \\
      NGC 0312  &     7.3  &    2.0   & 0.205 &  0.063  & 0.362  & 0.026 \\
      NGC 0484  &    10.4  &    2.0   & 0.222 &  0.059  & 0.155  & 0.022 \\
      NGC 0596  &     2.8  &    0.4   & 0.360 &  0.035  & 0.175  & 0.015 \\
      NGC 0636  &     4.7  &    4.1   & 0.360 &  0.188  & 0.225  & 0.056 \\
      NGC 0731  &     3.1  &    0.7   & 0.351 &  0.066  & 0.147  & 0.026 \\
      NGC 1052  &    21.7  &    1.6   & 0.222 &  0.025  & 0.390  & 0.007 \\
      NGC 1172  &     7.0  &    2.3   & 0.077 &  0.057  & 0.088  & 0.031 \\
      NGC 1316  &     3.2  &    0.2   & 0.338 &  0.023  & 0.150  & 0.009 \\
      NGC 1340  &     2.1  &    0.1   & 0.490 &  0.056  & 0.200  & 0.014 \\
      NGC 1375  &     2.6  &    3.1   & 0.011 &  0.177  &-0.044  & 0.133 \\
      NGC 1395  &     7.6  &    1.4   & 0.439 &  0.033  & 0.353  & 0.013 \\
      NGC 1400  &    15.0  &    1.6   & 0.226 &  0.032  & 0.338  & 0.010 \\
      NGC 1407  &     7.4  &    1.8   & 0.379 &  0.044  & 0.300  & 0.019 \\
      NGC 1411  &     9.1  &    0.9   & 0.042 &  0.026  & 0.163  & 0.015 \\
      NGC 1549  &     6.1  &    0.6   & 0.283 &  0.020  & 0.237  & 0.009 \\
      NGC 1553  &     4.7  &    0.3   & 0.370 &  0.017  & 0.188  & 0.009 \\
      NGC 2434  &     5.5  &    2.4   & 0.167 &  0.090  & 0.066  & 0.038 \\
      NGC 2717  &     3.7  &    2.8   & 0.460 &  0.128  & 0.300  & 0.042 \\
      NGC 2891  &     2.0  &    0.8   & 0.113 &  0.099  & 0.163  & 0.057 \\
      NGC 2986  &    11.2  &    1.9   & 0.347 &  0.045  & 0.250  & 0.014 \\
      NGC 3078  &    22.3  &    4.2   & 0.228 &  0.055  & 0.250  & 0.016 \\
      NGC 3224  &     1.9  &    0.7   & 0.554 &  0.139  & 0.166  & 0.041 \\
      NGC 3923  &     3.3  &    0.8   & 0.623 &  0.070  & 0.308  & 0.022 \\
      NGC 4786  &    10.4  &    2.4   & 0.237 &  0.063  & 0.213  & 0.024 \\
      NGC 5087  &     7.3  &    2.3   & 0.523 &  0.068  & 0.323  & 0.026 \\
      NGC 5516  &    11.2  &    2.1   & 0.380 &  0.046  & 0.131  & 0.017 \\
      NGC 5791  &     6.8  &    3.9   & 0.530 &  0.124  & 0.199  & 0.045 \\
      NGC 5903  &     8.8  &    2.4   & 0.264 &  0.075  & 0.186  & 0.029 \\
      NGC 6305  &     3.7  &    1.6   & 0.322 &  0.092  & 0.188  & 0.035 \\
      NGC 6958  &    11.3  &    1.0   & 0.015 &  0.027  & 0.097  & 0.013 \\
      NGC 7041  &     4.6  &    0.3   & 0.373 &  0.025  & 0.350  & 0.010 \\
      NGC 7185  &     2.3  &    0.9   & 0.145 &  0.100  & 0.038  & 0.047 \\
      NGC 7365  &     3.1  &    3.4   & 0.090 &  0.176  & 0.095  & 0.104 \\
      NGC 7796  &    11.8  &    1.9   & 0.248 &  0.050  & 0.344  & 0.019 \\
\enddata 
\tablecomments{The stellar population parameters age, total
metallicity [\ZH], and [\aFe] abundance ratio are derived from \Hb,
\Mgb, and \Fe\ (Table~\ref{tab:sample}) using SSP models of
\citet{TMB03a} and \citet{TMK04}. Cols.~3, 5, 7 give 1-$\sigma$
errors. The latter are estimated via Monte Carlo simulations (see
text).}
\end{deluxetable}

\clearpage
\section{The local abundance pattern at high metallicities}
\label{sec:bias}
In a recent study, \citet{Proetal04} argue that the correlation of
\aFe\ with velocity dispersion is an 'artefact of incomplete
calibration (of the models) to the Lick system'. The basis for their
conclusion is the recent finding of \citet{BFL04}, that the Milky Way
disk stars with super-solar metallicities do not have solar but
sub-solar O/Fe ratios, O/Fe decreasing with increasing Fe/H \citep[see
also][]{PBG03,Prietal04}. Stellar population models, being calibrated
with these stars, are then likely to reflect sub-solar O/Fe ratios at
high metallicities.  If the element Oxygen is taken as representative
for all $\alpha$-elements, this would obviously lead to an
overestimation of the O/Fe ratio (or \aFe\ ratio) in stellar
populations with super-solar metallicities like early-type
galaxies. \citet{Proetal04} show that the correction for this effect
makes the \aFe-$sigma$ relation disappear.

However, there is a further very important (and somewhat puzzling)
point shown in \citet{BFL04}. All the other $\alpha$-elements, Mg, Si,
Al, Na, Ca, and Ti, do not show this pattern, but level-off at a solar
X/Fe ratio in stars with super-solar metallicities \citep{Fu98,BFL03}.
Only the element oxygen (and maybe also Carbon, \citealt*{TEG04})
sticks out of this homogeneous behavior.  This differential pattern
between Oxygen and the other $\alpha$-elements cannot be neglected in
models taking the local abundance pattern at high metallicities into
account. This is particularly important in the context of this work,
as the \aFe\ ratios presented here are derived from the absorption
line indices \Mgb\ and \Fe, which are sensitive mostly to the elements
Fe and Mg, but {\em not} O \citep{TB95,KMT05}.

We investigated this issue constructing two additional flavors of
stellar population models that account for the local abundance pattern
at high metallicity. One in which Oxygen is taken as representative
for all $\alpha$-elements assuming these elements to be enriched in
lockstep (like in \citealt{Proetal04}), and one in which Oxygen is
detached from the rest of the $\alpha$-elements as discussed
above. The decrease of \OFe\ with metallicity is adopted directly from
\citet{BFL04} and can be parameterized as \citep{Proetal04}:
\begin{equation}
\label{eqn:bias}
[\OFe]=-0.5\ [\FeH]\ \ {\rm for}\ [\FeH]>0\ .
\end{equation}

\subsection{Oxygen and $\alpha$-elements in lockstep}
\begin{figure*}
\centering\includegraphics[width=0.7\linewidth]{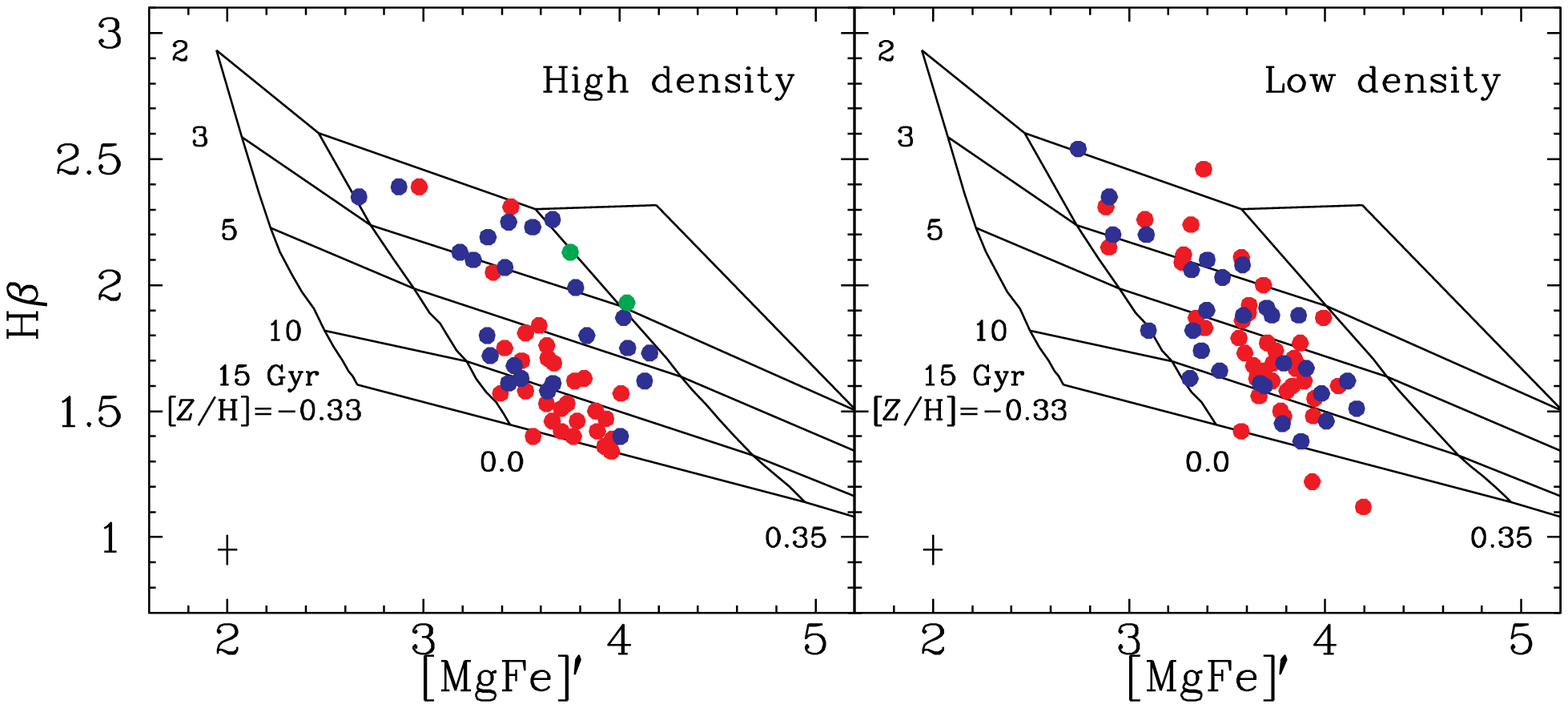}
\centering\includegraphics[width=0.7\linewidth]{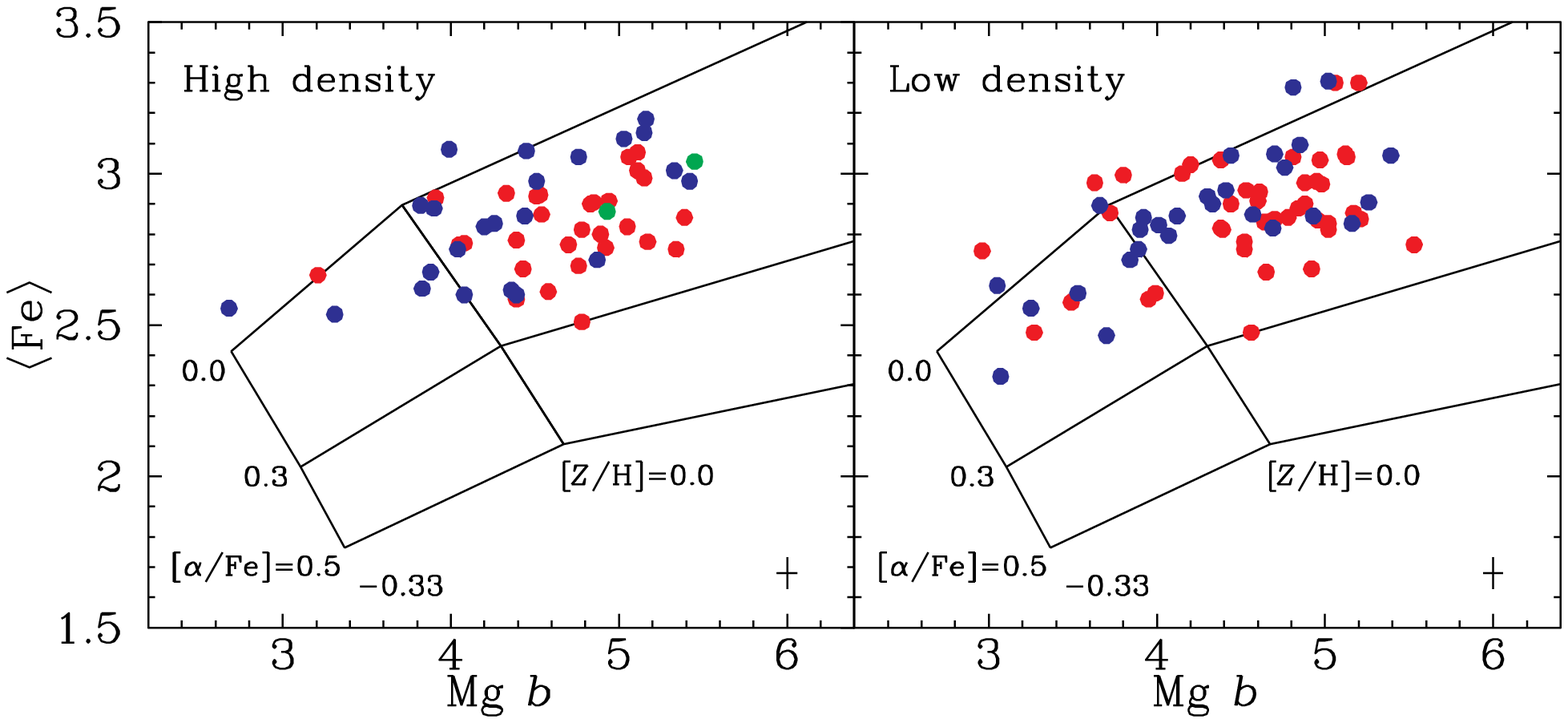}
\caption{Lick indices \MgFe\ versus \Hb\ (top panels) and \Mgb\ versus
\Fe\ (bottom panels). Symbols like in Fig.~\ref{fig:datagrid}.  SSP
models with the metallicities $[\ZH]=0.0,0.35,0.67$ are plotted for
different ages $t=2,3,5,10,15$~Gyr at fixed \aFe\ ratio ($[\aFe]=0$)
(top panels) and different \aFe\ ratios $[\aFe]=0.0,\ 0.3,\ 0.5$ at
fixed age ($t=12$~Gyr) (bottom panels) as indicated by the
labels. Different from Fig.~\ref{fig:datagrid}, the models are
corrected for the decrease of \OFe\ with increasing metallicity at
super-solar metallicity in the solar neighborhood assuming oxygen and
the other $\alpha$-elements being lumped together.}
\label{fig:biasgrid1}
\end{figure*}
The resulting model in the index-index planes, assuming oxygen as a
representative for all $\alpha$-elements, is shown in
Fig.~\ref{fig:biasgrid1} (equivalent to Fig.~\ref{fig:datagrid}).

The effect is striking. As oxygen dominates metallicity, the
correction for the sub-solar \OFe\ of the calibrating stars at high
metallicities has the effect that the models are practically 'labeled'
with lower metallicities. As a consequence, stronger metal indices at
a given metallicity are predicted, which leads to the derivation of
somewhat lower (but still super-solar) metallicities (top panel of
Fig.~\ref{fig:biasgrid1}). The derived ages, instead, remain the
unchanged. Even more dramatic is the modification of the model in the
\Mgb-\Fe\ plane (bottom panel). At super-solar metallicity, the \OFe\
(hence \aFe\ in this case) of the original model based on the local
abundance pattern has to be increased in order to obtain solar-scaled
abundance ratios. This is achieved mainly by decreasing the abundance
of iron, so that the iron indices get weaker and the \Mgb\ index get
stronger (due to the inverse response to Fe abundance, see TMB for
details). The model slope in this diagram is now close to the slope of
the data, which clearly impacts on the derived \aFe-$\sigma$ relation.

\begin{figure}
\centering\includegraphics[width=0.5\linewidth]{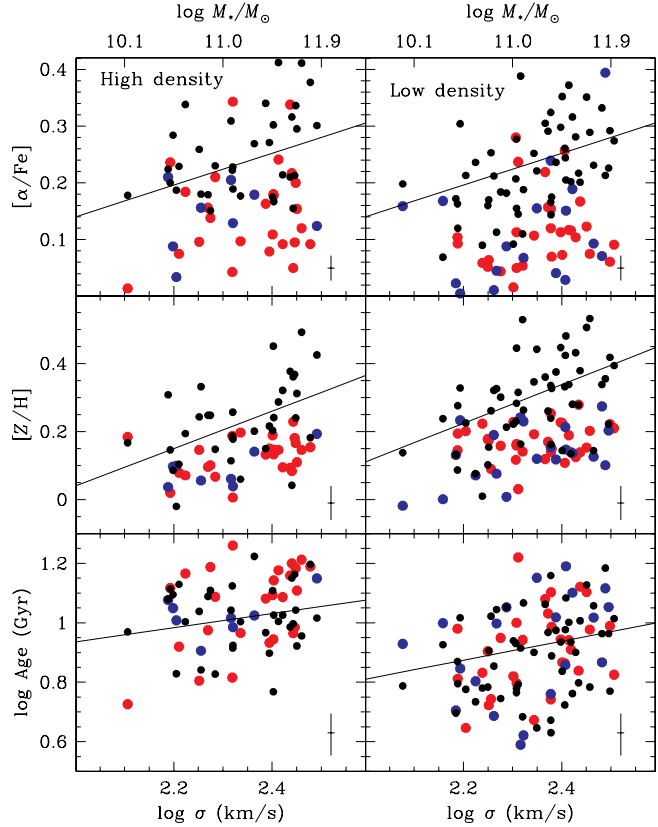}
\caption{Stellar population parameters as a functions of velocity
dispersion (measured within $1/10\ r_e$) and stellar mass (upper
x-axis) for the 'old' subpopulation (see Fig.~\ref{fig:popsall}).
Ages and element abundances are derived with the SSP model of
Fig.~\ref{fig:biasgrid1}. Black points are Monte Carlo simulations
asuming Eqn.~\ref{eqn:relations} (solid lines), which reproduce the
data in Fig.~\ref{fig:popsall}.}
\label{fig:popsall2}
\end{figure}
For a better illustration, the stellar population parameters of the
galaxy sample investigated here obtained with this model are plotted
in Fig.~\ref{fig:popsall2} (the equivalent to
Fig.~\ref{fig:popsall}). The relationships of Eqn.~\ref{eqn:relations}
plus the Monte Carlo realizations of Fig.~\ref{fig:popsall} are also
shown for comparison. The \aFe\ ratios are lower by about $0.1\;$dex,
and in particular the \aFe-$\sigma$ relation disappears. Metallicities
are also lower by about $0.1\;$dex, and the \ZH-$\sigma$ relationship
gets significantly tighter. These findings are in agreement with the
conclusions of \citet{Proetal04}. Note that the ages do not change.

\subsection{Oxygen and $\alpha$-elements separated}
\begin{figure*}
\centering\includegraphics[width=0.7\linewidth]{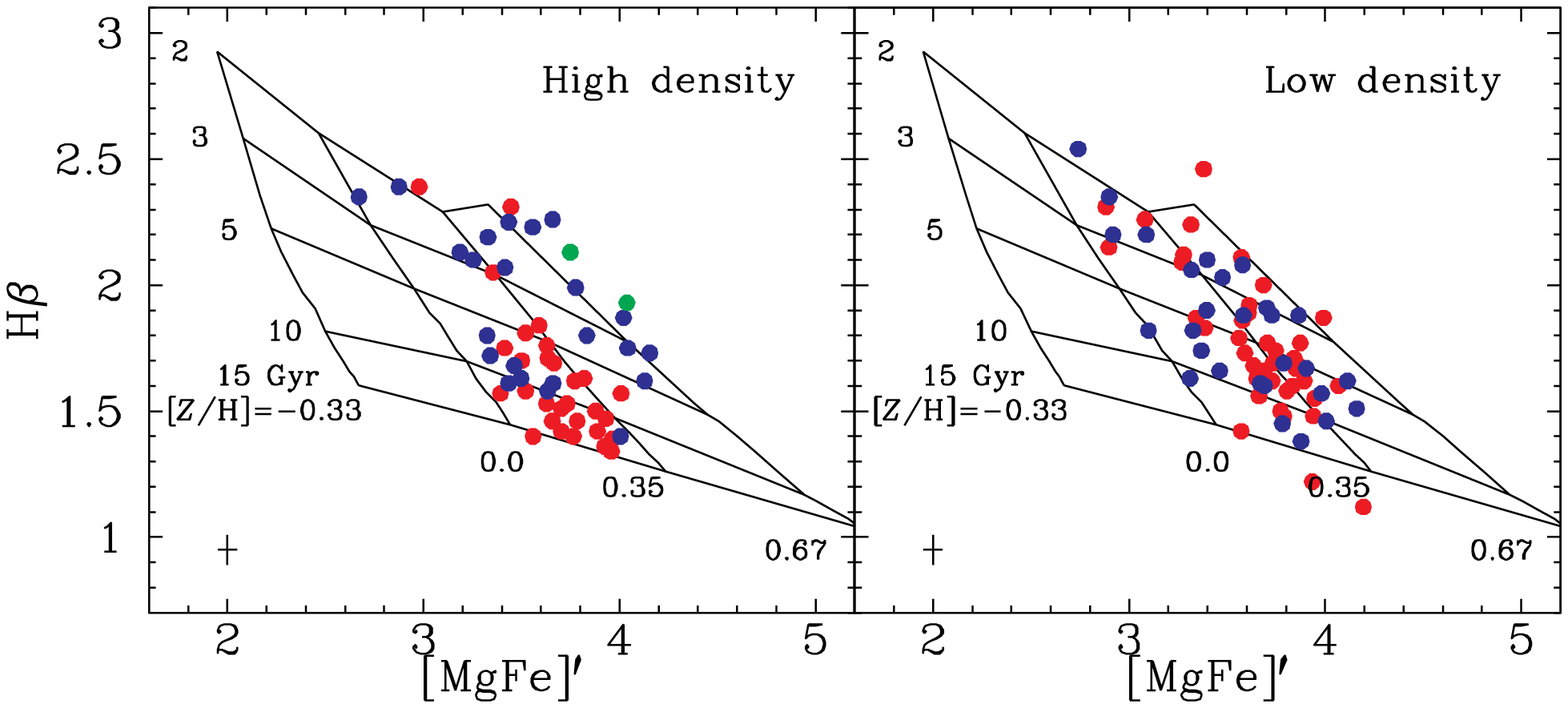}
\centering\includegraphics[width=0.7\linewidth]{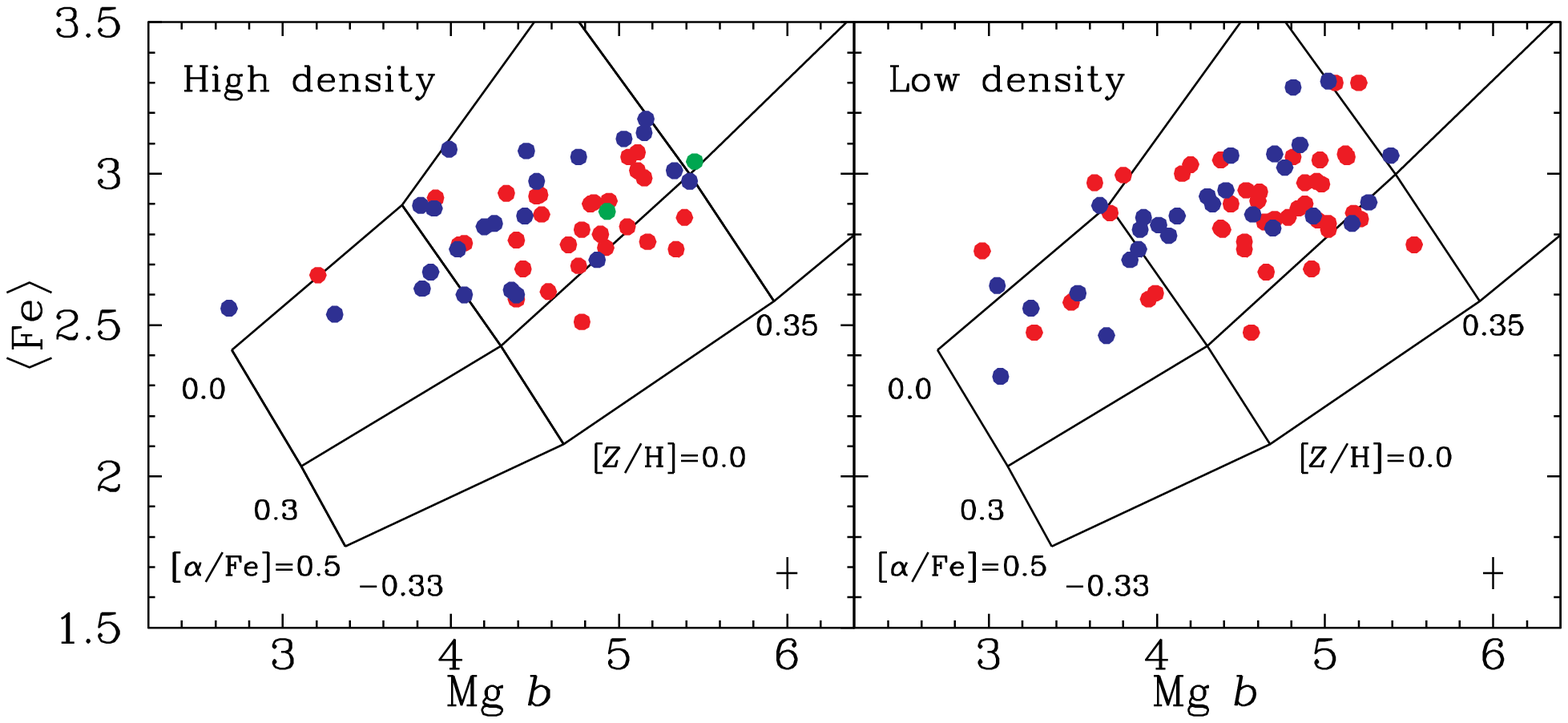}
\caption{Lick indices \MgFe\ versus \Hb\ (top panels) and \Mgb\ versus
\Fe\ (bottom panels). Symbols like in Fig.~\ref{fig:datagrid}.  SSP
models with the metallicities $[\ZH]=0.0,0.35,0.67$ are plotted for
different ages $t=2,3,5,10,15$~Gyr at fixed \aFe\ ratio ($[\aFe]=0$)
(top panels) and different \aFe\ ratios $[\aFe]=0.0,\ 0.3,\ 0.5$ at
fixed age ($t=12$~Gyr) (bottom panels) as indicated by the
labels. Different from Fig.~\ref{fig:datagrid}, the models are
corrected for the decrease of \OFe\ with increasing metallicity at
super-solar metallicity in the solar neighborhood assuming oxygen and
the other $\alpha$-elements being separated.}
\label{fig:biasgrid2}
\end{figure*}

In this second flavor of the model we account for the fact that the
other $\alpha$-elements (in particular magnesium) actually do not
follow the decrease of the local \OFe\ ratio with increasing
metallicity at super-solar metallicities \citep{BFL04}, hence
[Ne,Mg,Si,S,Ar,Ca,Ti/Fe]$= 0$ for $[\FeH]\geq 0$. The impact on the
index-index diagrams is shown in Fig.~\ref{fig:biasgrid2}. The model
is almost indistinguishable from the original model used for the
analysis of this paper (Fig.~\ref{fig:datagrid}) and very different
from the model in which oxygen and the other $\alpha$-elements are
lumped together (Fig.~\ref{fig:biasgrid1}). The reason is the
following. Oxygen contributes almost half, and the rest of the
$\alpha$-elements about one quarter to total metallicity. A variation
of the \aFe\ ratio {\em including oxygen} is therefore obtained mainly
by a modification of Fe abundance \citep[][and TMB]{Traetal00a}. 

\begin{figure}
\centering\includegraphics[width=0.5\linewidth]{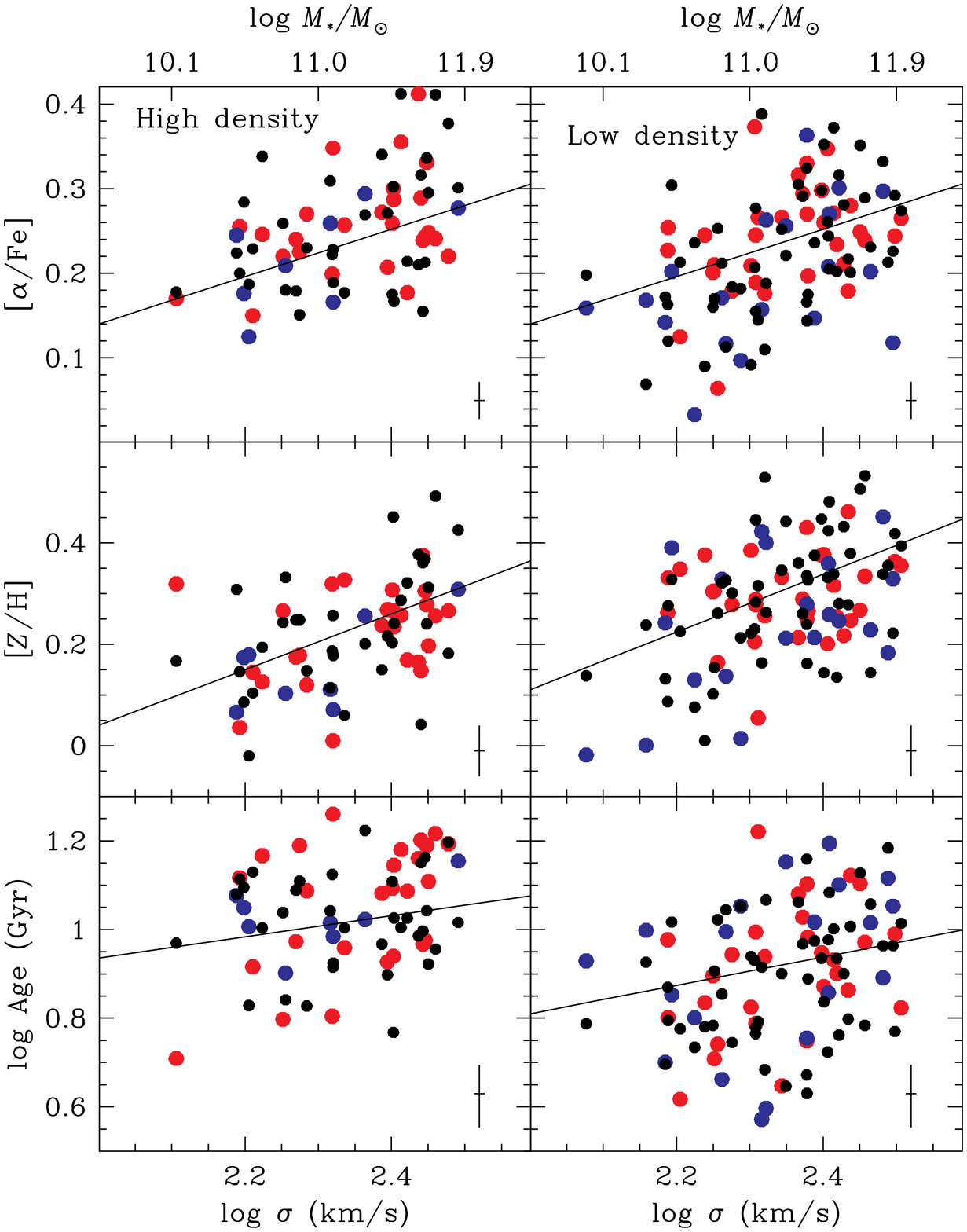}
\caption{Stellar population parameters as a functions of velocity
dispersion (measured within $1/10\ r_e$) and stellar mass (upper
x-axis) for the 'old' subpopulation (see Fig.~\ref{fig:popsall}).
Ages and element abundances are derived with the SSP model of
Fig.~\ref{fig:biasgrid2}. Black points are Monte Carlo simulations
asuming Eqn.~\ref{eqn:relations} (solid lines), which reproduce the
data in Fig.~\ref{fig:popsall}. }
\label{fig:popsall3}
\end{figure}
The correction from the local sub-solar \OFe\ to solar element ratios
is therefore achieved by a decrease of Fe abundance by 40 per
cent. When oxygen is detached from the other $\alpha$-elements,
instead, the picture is very different. The required decrease of Fe
abundance is only 20 per cent, which reduces both the decrease of the
Fe indices and the increase of \Mgb\ (note that \Mgb\ responses
inversely to Fe abundance). On top of this, the abundances of the
other $\alpha$-elements (including magnesium) need to be reduced by
the same amount to maintain the solar \aFe\ ratio. This leads to a
further decrease of \Mgb\ and an increase of the Fe indices, the
latter being inversely coupled with Mg abundance
\citep{TB95,KMT05}. The effect is sufficient to bring the model back
to its original position. As a consequence, the stellar population
parameters derived with this model are excellent agreement with the
ones presented in this work as shown in Fig.~\ref{fig:popsall3}.

To conclude, when the non-solar O/Mg ratios of metal-rich stars in the
solar neighborhood are taken into account, the correction for the
local abundance pattern has no significant impact on the model. This
implies that the results of this paper are robust against this
problem. In particular, the \aFe-$\sigma$ relation found here is real
and not caused by calibration artefacts.

\end{document}